\documentclass[twocolumn, twocolappendix]{aastex7}
\usepackage{amsmath}
\usepackage{CJK}
\usepackage{hyperref}
\defcitealias{2024ApJ...961..167L}{I}    

\begin{document}
\begin{CJK*}{UTF8}{bsmi}

\title{Convection-Driven Multi-Scale Magnetic Fields Determine the Observed Solar-Disk Gamma Rays}

\shortauthors{Li et al.}
\shorttitle{Modeling Solar-Disk Gamma Rays in Multi-Scale Magnetic Fields}


\author[0000-0003-1671-3171, gname='Jung-Tsung', sname='Li']{Jung-Tsung Li (李融宗)}
\affiliation{Center for Cosmology and AstroParticle Physics, The Ohio State University, Columbus, OH 43210, USA}
\email[show]{li.12638@osu.edu}

\author[0000-0003-0204-8385, gname='Mahboubeh', sname='Asgari-Targhi']{Mahboubeh Asgari-Targhi}
\affiliation{Harvard-Smithsonian Center for Astrophysics, 60 Garden Street, Cambridge, MA 02138, USA}
\affiliation{Department of Earth and Planetary Science, The University of Tokyo, 7-3-1 Hongo, Bunkyo-ku, Tokyo 113-0033, Japan}
\email[show]{masgari-targhi@cfa.harvard.edu}

\author[0000-0002-0005-2631, gname='John', sname='Beacom']{John F. Beacom}
\affiliation{Center for Cosmology and AstroParticle Physics, The Ohio State University, Columbus, OH 43210, USA}
\affiliation{Department of Physics, The Ohio State University, Columbus, OH 43210, USA}
\affiliation{Department of Astronomy, The Ohio State University, Columbus, OH 43210, USA}
\email{beacom.7@osu.edu}

\author[0000-0002-8040-6785, gname='Annika', sname='Peter']{Annika H. G. Peter}
\affiliation{Center for Cosmology and AstroParticle Physics, The Ohio State University, Columbus, OH 43210, USA}
\affiliation{Department of Physics, The Ohio State University, Columbus, OH 43210, USA}
\affiliation{Department of Astronomy, The Ohio State University, Columbus, OH 43210, USA}
\email{peter.33@osu.edu}

\date{\today}


\begin{abstract}
The solar disk is a continuous source of GeV--TeV gamma rays. The emission is thought to originate from hadronic Galactic cosmic rays (GCRs) interacting with the gas in the photosphere and uppermost convection zone after being reflected by solar magnetic fields. Despite this general understanding, existing theoretical models have yet to match observational data. At the photosphere and the uppermost convection zone, granular convection drives a multi-scale magnetic field, forming a larger-scale filamentary structure while also generating turbulence-scale Alfv\'{e}n wave turbulence. Here, we demonstrate that the larger-scale filamentary field shapes the overall gamma-ray emission spectrum, and the Alfv\'{e}n wave turbulence is critical for further suppressing the gamma-ray emission spectrum below $\sim 100$~GeV. For a standard Alfv\'{e}n wave turbulence level, our model's predicted spectrum slope from 1~GeV to 1~TeV is in excellent agreement with observations from Fermi-LAT and HAWC, an important achievement. The predicted absolute flux is a factor of 2--5 lower than the observed data; we outline future directions to resolve this discrepancy. The key contribution of our work is providing a new theoretical framework for using solar disk gamma-ray observations to probe hadronic GCR transport in the lower solar atmosphere.
\end{abstract}


\section{INTRODUCTION} \label{sec: introduction}

The origin of continuous, GeV--TeV gamma-ray emission from the solar disk has drawn considerable attention, propelled both by the advent of new gamma-ray observatories over the last two decades and by the unique perspective these high-energy emissions provide into the transport of GCRs in the solar atmosphere. Data from the Fermi Large Area Telescope (Fermi-LAT), in particular, has allowed the detection of gamma rays over a broad energy range from~0.1 to~200~GeV, revealing a hard spectrum following $dN_\gamma/dE_\gamma \sim E_\gamma^{-2.2}$ \citep{2011ApJ...734..116A, 2016PhRvD..94b3004N, 2018PhRvD..98f3019T, 2018PhRvL.121m1103L, 2022PhRvD.105f3013L}, and earlier Energetic Gamma Ray Experiment Telescope (EGRET) data offered indications of this phenomenon \citep{2008A&A...480..847O}. Pushing to even higher energies, the High Altitude Water Cherenkov Observatory (HAWC) detects gamma rays at around 1~TeV but finds a much softer spectrum following $dN_\gamma/dE_\gamma \sim E_\gamma^{-3.6}$ \citep{2023PhRvL.131e1201A}. Intriguingly, data from both experiments report an anti-correlation between gamma-ray flux and the solar cycle, with a flux approximately double at all energies at solar minimum compared to solar maximum \citep{2011ApJ...734..116A, 2022PhRvD.105f3013L, 2023PhRvL.131e1201A}. Furthermore, Fermi-LAT data reveal a moderate spatial anisotropy across the solar surface, which also varies with the solar cycle \citep{2018PhRvL.121m1103L, 2024ApJ...962...52A, 2025ApJ...989L..16A}. This dependence on the solar cycle in both the overall flux and its spatial distribution implies that the emission occurs across the solar disk and is indirectly affected by global solar magnetic activity.

It is now widely accepted that for the Sun to continuously generate these high-energy gamma rays, it must be that hadronic GCRs---primarily protons ($p$) and helium nuclei ($\mathrm{He^{2+}}$)---reach the lower solar atmosphere, where they collide with dense solar gas (primarily hydrogen and helium atoms) \citep[e.g.,][]{1965RvGSP...3..319D, 1966JGR....71.5778P, 1973NPhS..242...59S, Hudson1989}. These collisions initiate hadronic interactions that produce neutral pions (and other mesons), which subsequently decay into gamma rays, making the solar surface a gamma-ray source. (In the following, we use ``$pp$ interactions'' to refer to all such hadronic interactions, including those from $p + \mathrm{He^{2+}}$ and $\mathrm{He^{2+}} + \mathrm{He^{2+}}$ collisions.) The basic interaction processes can be described by $p + p \to p + p + \pi^0$, followed by $\pi_0 \to \gamma + \gamma$. In order for the emitted gamma rays to not be fully absorbed by the solar gas, hadronic GCRs must be reflected by magnetic fields from inward to outward trajectories prior to interacting. Without magnetic fields, multi-GeV to TeV emission would be confined to the solar limb \citep{2017PhRvD..96b3015Z}, contrary to the Fermi-LAT observations revealing emission across the entire solar disk \citep{2018PhRvL.121m1103L, 2024ApJ...962...52A}. Thus, the central question is to understand which magnetic structures are responsible for reflecting GCRs at the atmospheric depths critical for triggering the hadronic interactions.

Previous efforts to model GCR reflection have taken two approaches. The first approach is to model the magnetic field structures in the lower solar atmosphere. A pioneering study by \citet{1991ApJ...382..652S} assumes a pressure equilibrium following $P\left(z\right) \propto B\left(z\right)^2$, where $P\left(z\right)$ and $B\left(z\right)$ are gas and magnetic pressures at the height $z$. Particles are reflected via the magnetic mirror effect, with trajectories determined by adiabatic invariance. Within the considered $E_\gamma$ range of 0.1 to 5~GeV, their predicted flux is~5 to~10 times lower than the later Fermi-LAT data. An extrapolation of their predicted flux to 1~TeV is more than one order of magnitude below the later HAWC data. Despite this mismatch, their basic assumption, that the magnetic mirror effects in and near the lower solar atmosphere are responsible for enabling GCR to illuminate the Sun in gamma rays, remains influential in current attempts to reconcile detailed theoretical models to data. To refine the conventional $P\propto B^2$ scaling law, \citet{2020MNRAS.491.4852H} employs \texttt{Bifrost} magnetohydrodynamic (MHD) simulations to propose that $P \propto B^{3.5}$ may more accurately reflect solar conditions. The implications of this revised scaling for the disk emission have not yet been developed. \citet{2024ApJ...973..118P} considers closed magnetic arcades of 20~Mm in horizontal size at the photosphere and incorporates phenomenological magnetic turbulence to account for pitch-angle scattering between GCRs and turbulent fields---an effect not addressed in \citet{1991ApJ...382..652S}. Their gamma-ray prediction under strong turbulence conditions ($\delta B \sim B_0$ at the correlation length scale where $\delta B$ and $B_0$ are the turbulent and mean field strengths, respectively) aligns more closely with the Fermi-LAT data, whereas their prediction under weak turbulence ($\delta B \ll B_0$) is three orders of magnitude lower. \citet{2025ApJ...989L...3P} proposes an open magnetic field model with a synthetic, isotropic braiding magnetic field over a constant mean field ($B_0 = 5$~G throughout the simulation domain). The maximum magnetic field strength reaches $\sim 100$~G at the photosphere. Their predicted gamma-ray spectrum slope is $dN_\gamma / dE_\gamma \sim E_\gamma^{-2.7}$, which is similar to that of the GCR spectrum and steeper than the observed gamma-ray data by Fermi-LAT. This could arise if the magnetic field structures effectively mirror the cosmic-ray spectrum without energy dependence. It would be interesting to explore their magnetic structures further, in particular their horizontal extent (42~Mm) at the photosphere, which seems to us to be in conflict with solar data.

The second approach to model GCR reflection is to construct the global, large-scale coronal magnetic field of \emph{the entire Sun} using, for example, the Potential Field Source Surface (PFSS) model. The PFSS model extrapolates line-of-sight photospheric magnetogram data (lower boundary) into the corona by solving the Laplace equation for the scalar magnetic potential and enforcing the magnetic field lines to become radial at the source surface (upper boundary), typically located at a heliocentric distance of about $2.5 \, R_\odot$ \citep{1969SoPh....6..442S, 1969SoPh....9..131A}. The simplicity of the PFSS approach has made it a common tool for modeling the coronal magnetic field in the context of solar gamma-ray emission \citep{2020PhRvD.101h3011M, 2024ChPhC..48d5101L}. These PFSS simulations for the solar gamma-ray emission model the entire solar coronal field. However, such a global scope requires a coarse simulation grid (with resolution $\gtrsim 10$~Mm), which limits the resolution of turbulence-scale magnetic fields in the photosphere and chromosphere. While the predictions from this approach show some agreement with Fermi-LAT data for $E_\gamma \lesssim 10$~GeV, they deviate at higher energies, particularly when compared to HAWC data at $1$~TeV. Furthermore, the temporal modulation predicted by this approach is inconsistent with observations.

Beyond these two approaches, several novel ideas have been proposed to specifically explain the HAWC's TeV gamma-ray signal. \citet{2020APh...11902440G} and \citet{2022ApJ...941...86G} use the Sun's cosmic-ray shadow data to estimate the fraction of GCRs absorbed by the Sun and the resulting gamma-ray yields. \citet{2023arXiv230517086B} instead attributes the TeV emission to thermal protons accelerated to multi-TeV energies by acoustic-like shocks in the chromosphere. Last, \citet{2025PhRvL.135l5201N} argues that kilo-Gauss horizontal fields at and beneath the photosphere in the internetwork region can reflect multi-TeV GCRs and reproduce the observed TeV signal.

Thus, while existing models investigate GCR reflection by solar magnetic fields and show partial agreement in their predictions, they disagree significantly on the relevant spatial scales of the magnetic fields and the atmospheric layers that govern the observed gamma-ray emission. Moreover, these models neglect the filamentary magnetic structures ($\sim \mathrm{few} \times 100$~km) known to permeate the photosphere. To resolve this divergence, it is essential to first identify the atmospheric layers where the hadronic interactions occur, followed by characterizing the spatial properties of magnetic fields within these layers.

\subsection{A Convective Photosphere}

We can estimate the hadronic cascading layers by considering the mean free path of proton GCRs in the solar gas and imposing the condition $\int \sigma_{pp} n_\mathrm{H}\left(z\right) dz \gtrsim 1$. (Note that particle gyro-motion is neglected in this simple estimation but is fully accounted for in our numerical simulations.) Here, $\sigma_{pp} \simeq 3 \times 10^{-26}~\mathrm{cm^2}$ is the total inelastic hadronic cross section for $pp$ interactions at the relevant GCR kinetic energies, and $n_\mathrm{H}\left(z\right)$ is the hydrogen number density of the gas at height $z$ from the photosphere base ($z=0$~km, where the optical depth $\tau_{5000}$ for $\text{5000~\AA}$ light is unity). At $z=0$~km, $n_\mathrm{H} \sim 10^{18} \, \mathrm{cm^{-3}}$, with the scale height of $\sim 100$~km above and $\sim 250$~km below~\citep{2008ApJS..175..229A}. Thus, the interaction height is estimated to span in the order of a few hundred~km above and below the photosphere base, covering the lower chromosphere, the photosphere, and the uppermost convection zone. In this work, we will show that the emission occurs primarily in the $z=-100$~km to $z=400$~km height range.

In this height range, the magnetic field properties are governed by convective granular flows in the uppermost convection zone, giving rise to two distinct scales of magnetic structures. \emph{On larger scales (with horizontal extent of a~few hundred~km):} The distribution of photospheric magnetic fields is filamentary, occupying only a small fraction of the surface area. Such structure is also known as ``small-scale'' magnetic field. This filamentary phenomenon occurs because magnetic fields are expelled to cellular edges by the convective granular flows in the uppermost convection zone. This process, known as flux expulsion, forms high-intensity, kilo-Gauss magnetic regions at the edges of granules, such as intergranular lanes and micropores \citep{1963ApJ...138..552P, 1966RSPSA.293..310W, 1981ApJ...243..945G}. Concurrently, granular flows constantly jostle each other, launching outward-directed Alfv\'{e}n waves into the photosphere and above. As these waves move outward, they encounter inward-directed Alfv\'{e}n waves reflected from the chromosphere and lower corona \citep[e.g.,][]{1958ApJ...127..459F, 1980JGR....85.1311H, 1989PhRvL..63.1807V, 1993A&A...270..304V, 2007JGRA..112.8102H}. \emph{On smaller, magnetic turbulence scales:} The resulting counter-propagating Alfv\'{e}n waves generate low-frequency virtual modes that couple to fluid eddy turbulence, leading to a cascade towards smaller perpendicular scales \citep{1964SvA.....7..566I, 1967PhFl...10.1417K}. Because GCRs are simultaneously guided by strong, filamentary magnetic structures and pitch-angle scattered by magnetic turbulence, a multi-scale treatment of the solar magnetic field is required to model their motion and subsequent gamma-ray emission.

Motivated by the filamentary nature of the photospheric magnetic field, \citet[hereafter Paper~\citetalias{2024ApJ...961..167L}]{2024ApJ...961..167L} modeled GCR reflection and subsequent gamma-ray emission in small-scale, filamentary magnetic structures (with horizontal extent of a~few hundred~km). The model consists of static, finite-sized flux tubes (representing individual constituents of a network element) and flux sheets (representing magnetic fields at intergranular lanes), each having a horizontal size of a few hundred km at the photosphere base. The model suggests that lower-energy gamma rays ($\lesssim 10$~GeV) originate mainly from network elements, while the higher-energy emission ($\gtrsim \mathrm{few} \times 10$~GeV) originates mainly from intergranular lanes. Notably, the predicted spectrum slope at 1~TeV aligns well with HAWC data, providing strong support for the critical role of filamentary structures in determining the observed gamma-ray spectra. A key insight in Paper~\citetalias{2024ApJ...961..167L} is that the observed gamma-ray patterns likely reflect the combined influence of various filamentary, small-scale magnetic flux structures, not just one type.

Paper~\citetalias{2024ApJ...961..167L}, however, has three limitations. First, because its magnetic field was modeled only up to $z=1600$~km, the overlying network magnetic field was excluded, potentially leading to the neglect of its role in GCR reflection via the magnetic mirror effect. Second, the strong reflection of Alfv\'{e}n waves at the transition region ($z \sim \text{2--3}$~Mm) of the network field generates counter-propagating waves that cascade their energy into the chromosphere and lower corona, creating regions of significant turbulence \citep[e.g.,][]{2011ApJ...736....3V, 2012ApJ...746...81A}. GCRs en route to hadronic cascading layers must traverse these turbulent areas. Paper~\citetalias{2024ApJ...961..167L}, however, did not investigate their effect on GCR propagation. Third, while the observations from Fermi-LAT extend down to $E_\gamma = 0.1$~GeV, the predictions in Paper~\citetalias{2024ApJ...961..167L} were restricted to $E_\gamma \gtrsim 3.66$~GeV due to limitations in the hadronic interaction model from \citet{2006PhRvD..74c4018K}.

\subsection{A Multi-scale Magnetic Field Model}

Building upon Paper~\citetalias{2024ApJ...961..167L}, the present work develops a more comprehensive magnetic field model that incorporates (1) the network field, (2) the small-scale, filamentary magnetic field, and (3) the Alfv\'{e}n wave turbulence. The objective is to investigate how multi-scale magnetic fields influence GCR propagation and the resulting gamma-ray emission. This work considers three distinct magnetic structures, each with a different scale:
\begin{enumerate}
    \item A network field structure from $z=700$~km to the corona. Each of the structure has a horizontal extent of $\sim \mathrm{few~Mm}$ at $z \sim 4$~Mm. Above the merging height of $z = 2-4$~Mm, nearby networks merge together, filling the entire volume with magnetic fields.
    \item The small-scale, filamentary magnetic field in the uppermost convection zone and photosphere layer, with horizontal extent of $\sim \mathrm{few \times 100~km}$ at $z=0$~km.
    \item The Alfv\'{e}n wave turbulence, for which the largest relevant spatial scales at $\sim 30$~km at $z=0$~km and cascading down to smaller and smaller spatial scales.
\end{enumerate}

Our methodology for analyzing the magnetic field separates it into two components: a turbulent part and a mean field part. For the turbulent part, we utilize a three-dimensional (3D) MHD model to simulate the generation, propagation, and dissipation of Alfv\'{e}n waves along seven open magnetic field lines. The Alfv\'{e}n wave turbulence driven by nonlinear interactions between counter-propagating Alfv\'{e}n waves is described using the ``reduced MHD'' (RMHD) approximation, following the numerical methods of \citet{2011ApJ...736....3V} and \citet{2012ApJ...746...81A}. For the mean field part, we model the magnetostatic 3D spatial structures of the network field and flux tube for seven open field lines from the RMHD analysis, along with the flux sheet, using the quasi-linear Grad-Shafranov equations described in \citet{1986A&A...170..126S}. These solutions yield a characteristic ``inverted wine bottle''-like geometry, where magnetic flux structures are narrowed at lower heights by higher ambient gas pressure and expand laterally as this pressure diminishes with increasing altitude (e.g., Paper~\citetalias{2024ApJ...961..167L}). The simulated Alfv\'{e}n wave turbulence is then superimposed onto the magnetostatic fields to construct the final, composite magnetic field solution.

We perform simulations of hadronic GCR propagation and gamma-ray emission in a height range from $z=-200$~km to $z=12$~Mm. We adopt the hadronic interaction model from \citet{2014PhRvD..90l3014K} to enable the gamma-ray prediction down to $E_\gamma = 0.1$~GeV. Our paper investigates how magnetic conditions along seven solar open field lines, categorized by their proximity near or far from active regions, influence the observed gamma-ray flux and spectrum slope. We further quantify the modulation of gamma-ray flux by varying Alfv\'{e}n wave turbulence levels. These findings offer new insights into GCR transport in the lower solar atmosphere and present a pathway for developing new probes of photospheric magnetic fields.

Last, gamma-ray emission from solar flares is episodic \citep[e.g.,][]{2012ApJ...745..144A, 2014ApJ...787...15A, 2014ApJ...789...20A, 2021ApJS..252...13A, 2015ApJ...805L..15P, 2024A&A...683A.208P, 2018ApJ...865L...7O, 2018ApJ...869..182S} and has been excluded from long-term gamma-ray observational data sets \citep{2016PhRvD..94b3004N, 2022PhRvD.105f3013L}. Thus, solar flares do not contribute to the continuous gamma-ray emission investigated herein.

This paper is organized as follows. In Section~\ref{sec: model setup}, we describe our model setup for multi-scale magnetic fields. In Section~\ref{sec: Alfven wave model}, we simulate the Alfv\'{e}n wave turbulence. In Section~\ref{sec: magnetostatic model}, we present the magnetostatic solutions. Our simulation setup for GCR propagation and gamma-ray emission is outlined in Section~\ref{sec: simulation setup}. Numerical results are shown in Section~\ref{sec: results}, followed by the Discussion in Section~\ref{sec: discussion}. Our conclusions are provided in Section~\ref{sec: conclusion}.

\begin{figure*}[t]
   \centering
   \includegraphics[width=1.7\columnwidth]{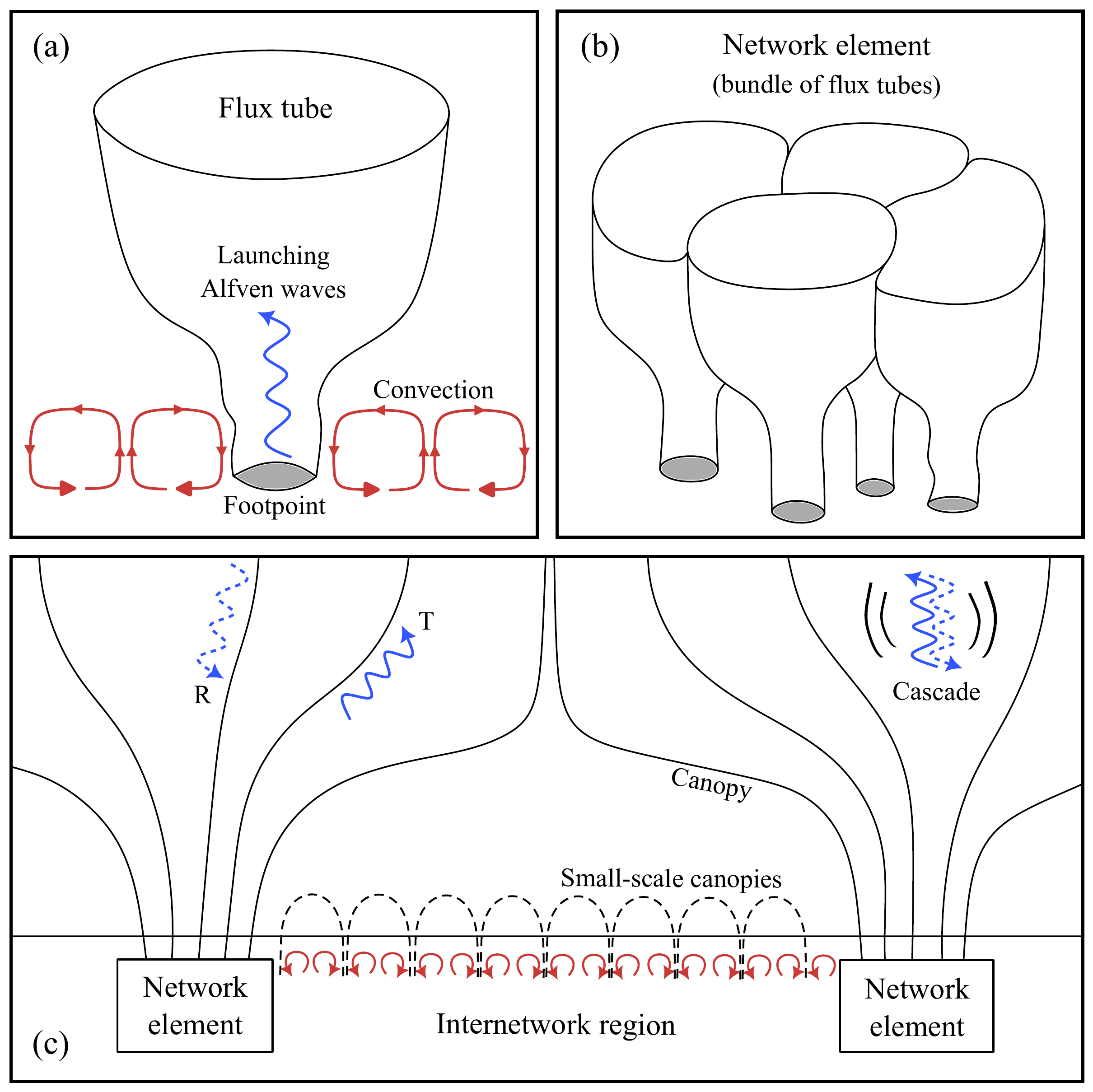}
   \caption{Schematic (not to scale) illustrating open magnetic network structure and Alfv\'{e}n wave turbulence. (a) and (b) are partially inspired by figures in \citet{2011ApJ...736....3V}, \citet{2005ApJS..156..265C}. (c) is inspired by \citet{2009SSRv..144..317W}. Panels (a)--(c) depict progressively larger viewing areas. (a) A flux tube emerges from the photosphere, with kilo-Gauss vertical magnetic fields at the footpoint (gray patch). Convective motions from granules (red arrows) in the uppermost convection zone and photosphere jostle the flux tubes, launching Alfv\'{e}n waves upward. (b) Individual flux tubes expand laterally with increasing height. At heights of 600--800~km, a bundle of tubes merge to form a network element. (c) Formed by field lines (solid lines) from network elements, a magnetic network further expands laterally with height and merges with adjacent networks. Portions of the  Alfv\'{e}n waves launched from the footpoints transmit (T) through this height range, while others are reflected (R). These counter-propagating Alfv\'{e}n waves generate an energy cascade. The outer edge of each network is a canopy that separates the magnetic network from the weak-field sub-canopy domain. The internetwork region, located between network elements, contains closed magnetic loops (``small-scale canopies'') that enclose the granules.}
   \label{fig:schematic}
\end{figure*}


\section{Model for Gamma-Ray Emission in Open Field Lines} \label{sec: model setup}

In this section, we present a simplified illustration of the multi-scale magnetic geometry of open field lines, which we hypothesize is critical to the formation of the observed gamma-ray pattern. We justify the vertical height range selected for our magnetic field model and review solar modulation.

\subsection{Simplified Picture of Multi-scale Open Magnetic Fields}

Figure~\ref{fig:schematic} illustrates the multi-scale geometry of open network field lines. The fundamental component of this open network is the magnetic flux tube (Figure~\ref{fig:schematic}(a)) at the uppermost convection zone and photosphere. These flux tubes typically have a field strength of 1--2~kilo-Gauss and a diameter of about 200~km at their footpoints (gray patch) at the photospheric base. Individual flux tubes are oriented predominantly vertically and maintain pressure equilibrium with the ambient gas. As gas pressure decreases with increasing height, each flux tube expands laterally. The first merger occurs at heights of about 500--700~km, where multiple flux tubes coalesce to form a network element (Figure~\ref{fig:schematic}(b)). Above $z=700$~km, each network element acts as a source for the monolithic magnetic network field, which can be modeled as a thicker magnetic flux structure (Figure~\ref{fig:schematic}(c)). Within this framework, the network field can be conceptualized as a bundle of magnetic field lines, where the origin of each line traces to a flux tube in the uppermost convection zone and photosphere. This bundle of lines continues its lateral expansion with height, leading to a second merging event at the height of 2--4~Mm where adjacent network field structures coalesce. Between $z=700$~km and 2--4~Mm, the outer edge of each network forms a magnetic canopy separating the magnetic network from the weak-field sub-canopy domain. Above the second merger height, the merged network magnetic fields fill the entire available volume and expand predominantly vertically.

Convective motions from granules (red arrows in Figure~\ref{fig:schematic}(a)) in the uppermost convection zone and photosphere jostle embedded magnetic flux tubes, thereby generating the upward-propagating Alfv\'{e}n waves. Due to the positive radial gradient of the Alfv\'{e}n speed in the upward direction, these waves undergo reflection as they ascend. The reflection is strongest at the chromosphere-corona transition region, where the Alfv\'{e}n speed abruptly increases by one order of magnitude \citep{2011ApJ...736....3V}. Consequently, while a portion of these Alfv\'{e}n waves transmits (labeled ``T''), the remainder is reflected (``R''; Figure~\ref{fig:schematic}(c)). These counter-propagating Alfv\'{e}n waves drive nonlinear wave-wave interaction, cascading magnetic energy to smaller perpendicular scales.

In Figure~\ref{fig:schematic}(c), the area between network elements is the internetwork region that hosts closed magnetic loops (``small-scale canopies''; dashed arcs) that typically enclose granules with their two polarities rooted in the adjacent intergranular lanes \citep{2005ESASP.596E..65S, 2006ASPC..354..345S, 2009SSRv..144..317W}. Granular convection is the primary driver of their formation. First, this convection drives flux expulsion, sweeping magnetic fields from granule interiors into intergranular lanes and creating vertically oriented magnetic sheets (two ends of the arc). The same convective flows also transport a portion of this magnetic flux upward to heights of about 500--600~km via the convective overshoot, forming the predominantly horizontal magnetic fields (top of the arc) that overlie the granules. Due to the absence of direct observations of the magnetic field structure below the photosphere base, our understanding of the corresponding intergranular sheet structures relies on state-of-the-art numerical magnetoconvection simulations. These simulations suggest that coherent intergranular sheets typically extend 500--1000~km beneath the surface before being disrupted by turbulent flows \citep[e.g.,][]{2005A&A...429..335V, 2005ESASP.596E..65S, 2006ASPC..354..345S, 2009A&A...504..595Y}.

\subsection{Model Height Range}

In our numerical computation of GCR propagation and gamma-ray emission, the upper boundary of the computational model is set at $z=12$~Mm. Below this height, the network field structure is modeled as a magnetostatic structure, following the approach of \citet{2005ApJS..156..265C}. At this upper boundary, we inject hadronic GCRs isotropically to trace their trajectories towards the solar surface. The propagation of GCRs above this height can be described as a solar modulation effect, which is addressed in the next subsection.

The lower boundary of our computational model is set at $z=-200$~km. This choice for the lower boundary is based on our numerical simulations of gamma-ray production (as we show in the gamma-ray emission height result in Figure~\ref{fig:f2_emitting_angle_height}), which indicate that over 95\% of the detectable gamma rays originate from the atmospheric layer between $z=-100$~km and $z=400$~km. Our tests using deeper lower boundaries of $z=-300$~km and $z=-400$~km confirmed that the flux and the height of the emission region remained unchanged. As a result, placing the boundary at $z=-200$~km is therefore sufficient to include the entire relevant production region. Furthermore, by setting the boundary at this depth, we can also reasonably assume that turbulent flows known to exist at greater depths do not significantly alter the magnetic field structure within this primary gamma-ray production region.

\subsection{Solar Modulation}

As GCRs in the interplanetary space and corona propagate toward the Sun, their flux decreases as they scatter off the magnetic turbulence in a radially expanding solar wind. This phenomenon is known as solar modulation \citep{1965P&SS...13....9P, 1967ApJ...149L.115G, 1968ApJ...154.1011G}. At the heliocentric distance of 1~au, numerous experiments have provided measurements of proton GCR spectra \citep[e.g.,][]{2013ApJ...765...91A, 2015PhRvL.114q1103A, 2021PhRvL.127A1102A, 2022ApJ...940..107C}, which include the solar modulation effects from the heliopause to 1~au. Because there have not yet been direct measurements of proton and helium GCR flux spectra near the Sun, GCR transport from 1~au to near the solar surface requires theoretical modeling. In \citet{2022ApJ...937...27L}, we utilized recent magnetic power spectrum measurements from the Parker Solar Probe within an improved force-field model to evaluate the reduction in GCR intensity. We find that the proton GCR intensity reduction from 1~au to 0.1~au is estimated to be $\lesssim 15\%$ for protons in the kinetic energy ranges of $0.1~\mathrm{GeV} \lesssim E_p^\mathrm{k} \lesssim 1~\mathrm{GeV}$. For higher-energy GCRs of $E_p^\mathrm{k} \geq 10$~GeV, the effect is less than $\mathcal{O} \left( 1\% \right)$. The observational data of the radial gradient from Helios~1 and~2 show an even smaller GCR intensity reduction of only $2 \pm 2.5 \%$ reduction from 1~au to 0.3~au for $E_p^\mathrm{k}$ in the ranges of $0.25~\mathrm{GeV} \leq E_p^\mathrm{k} \leq 0.7~\mathrm{GeV}$ \citep{2019A&A...625A.153M}. Furthermore, measurements from Mercury Surface, Space Environment, Geochemistry and Ranging (MESSENGER) between 0.3~au and 0.4~au show a proton GCR radial gradient of less than 10\% per au for energies $E_p^\mathrm{k} > 125$~MeV \citep{2016JGRA..121.7398L}. Based on the above theoretical estimates and the observational data, the solar modulation from 1~au to 0.1~au is likely a small effect and hence neglected.

From 0.1~au into the lower coronal region at $z=12$~Mm, the GCR transport is poorly constrained. The absence of direct GCR radial gradient measurements in this region also prevents a quantitative assessment of the solar modulation effect. In this paper, we do not attempt a direct, three-dimensional simulation of GCR trajectories from 1~au to $z=12$~Mm, as this approach is theoretically and computationally complex and beyond the scope of the present work. The primary challenge lies in tracking a statistically significant number of particles over an enormous spatial domain to accurately capture the diffusive motion that arises from their repeated pitch-angle scattering off the magnetic turbulence in the expanding solar wind. In addition to GCR scattering by magnetic turbulence, magnetic mirroring in the coronal magnetic field may induce anisotropies in the GCR distribution. Such anisotropies imply that a GCR angular distribution dominated by horizontally directed particles at $z=12$~Mm would produce a lower gamma-ray flux, whereas a distribution with more vertically directed particles would yield a higher gamma-ray flux. However, the magnitude of this anisotropy has not yet been quantified.

Therefore, to mitigate the uncertainties inherent in GCR transport models for this region, we neglect the effects of solar modulation and anisotropy between 1~au and our injection height of $z=12$~Mm. Consequently, we use the GCR spectra measured at 1~au as the direct input for injection into the network magnetic field at $z=12$~Mm. The proton GCR spectrum is constructed using AMS-02 data \citep{2015PhRvL.114q1103A} for the range $0.5~{\rm GeV} \leq E_p^{\rm k} \leq 1.46\times 10^3~{\rm GeV}$, and from CREAM \citep{2022ApJ...940..107C} for $2.10\times 10^3~{\rm GeV} \leq E_p^{\rm k} \leq 5.33\times 10^5~{\rm GeV}$. The gamma-ray contribution from helium and heavier nuclei in GCR and solar gas is accounted for using a standard nuclear enhancement factor, which we describe in Section~\ref{sec:gamma-emission pp collision}.

Finally, we turn to a promising method for probing the solar modulation. Observations have demonstrated that gamma-ray emission from the solar halo could diagnose the electron GCR intensity within 1~au from the Sun \citep{2006ApJ...652L..65M, 2007Ap&SS.309..359O, 2008A&A...480..847O, 2021JCAP...04..004O, 2011ApJ...734..116A, 2022PhRvD.105f3013L, 2025PhRvD.112j3030L, 2025PhRvD.111l3011M, 2025ApJ...989L..16A}. The emission from the solar halo originates from the inverse-Compton scattering of electron GCRs on solar photons. The detection of such gamma-ray signals from the solar corona thus confirms that electron GCRs must penetrate deep into the inner heliosphere and corona. This technique shows a unique probe of solar modulation of regions inaccessible to \emph{in-situ} spacecraft, including the corona and high-latitude heliospheric regions. While the presence of these electron GCRs in the solar corona is confirmed, the precise degree of modulation they experience en route to the solar surface remains to be quantified \citep{2011ApJ...734..116A, 2025PhRvD.112j3030L}. Furthermore, the modulation of proton and helium GCRs differs from that of electron GCRs because of their different charge-to-mass ratios \citep{2022ApJ...937...27L}. Therefore, developing a methodology to determine the solar modulation of hadronic GCRs in the inner heliosphere using solar halo gamma-ray emission data remains a critical area for future research.

\begin{figure}[t]
    \centering
    \includegraphics[width=0.99 \columnwidth]{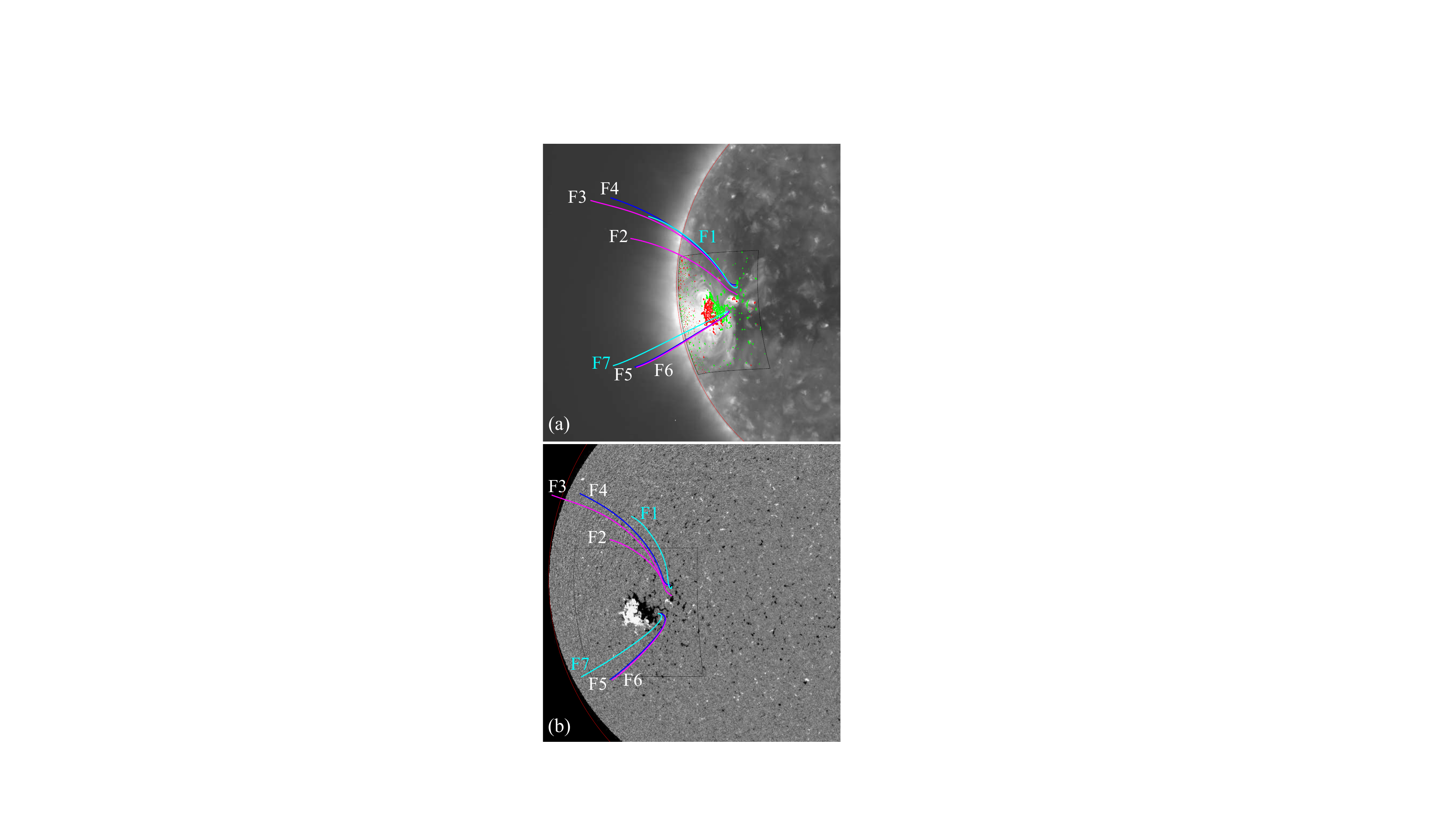}
    \caption{(a) Stereo-A HI image (195~{\AA}) of an active region close to the equator and to the East limb observed on 2007 December~09 at {06:35:30~UT}. Seven open network field lines (F1 to F7) are randomly selected using the potential field extrapolation. Different colors are chosen to aid visualization. Positive and negative polarities are indicated by red and green, respectively. (b) SOHO/MDI magnetogram observed on 2007 December~09 at {06:23:30~UT}. White and black represent positive and negative polarities, respectively. The line colors in both panels are chosen solely for visual clarity and do not carry any physical or categorical meaning.}
    \label{fig:fieldline_visualization}
\end{figure}


\section{Alfv\'{e}n Wave Turbulence Model In An Open Magnetic Field} \label{sec: Alfven wave model}

In this section, we first describe our selection of seven open magnetic field lines, four originating from a quiet region of the Sun and three from an active region. We then present the 3D MHD model developed to simulate the generation, propagation, and dissipation of Alfv\'{e}n waves along these chosen field lines and present the resulting findings.

\subsection{Magnetic Field Observations}

The photospheric region analyzed in this study is NOAA~10978, a bipolar structure that crossed the central meridian on 2007 December~09. We use the STEREO-A \citep{2007A&G....48c..12D} data taken around 06:35:30~UT on that day. Figure~\ref{fig:fieldline_visualization}(a) shows the magnetic flux distribution in the photosphere (red and green contours) overlaid on the 195~{\AA} image from the heliospheric imager (HI) \citep{2005AdSpR..36.1512H} on board STEREO-A. The flux distribution is derived from a line-of-sight magnetogram captured by the Michelson Doppler Imager (MDI) onboard the Solar and Heliospheric Observatory \citep[SOHO;][]{1995SoPh..162..129S} on 2007 December~09 at 06:23:30~UT. Sunspots are present in this region, and the HI image reveals coronal loops, fans, and open fields that extend from the sunspots and surrounding areas.

This active region is positioned near the equator and shows no evidence of twist or sigmoidal structures, making it suitable for modeling with the PFSS model. In a PFSS model, the magnetic field is assumed to be current-free, i.e., the volume current density $\boldsymbol{J}$ is zero. The model is constructed using the flux distribution from the MDI magnetogram and the SOLIS synoptic map corresponding to Carrington rotation 2064. The PFSS was computed using the Coronal Modeling System (CMS) software \citep{2004ApJ...612..519V, 2012ApJ...746...81A, 2013ApJ...773..111A, 2021ApJ...910..113A}, which enables a realistic model of the coronal magnetic field at low heights $z < 3 \, R_\odot$ \citep{2013ApJ...773..111A}. A selection of seven magnetic network field lines (F1--F7) traced through the PFSS model is shown in Figure~\ref{fig:fieldline_visualization}, superimposed on (a) the 195~{\AA} HI image and (b) the MDI magnetogram. As visualized in both images, F1--F4 originate from quieter patches of the Sun, whereas F5--F7 are rooted near an active region. This specific photospheric patch was selected as an initial case study. By including field lines from both quiet and active areas, we aim to understand how different local magnetic environments affect the gamma-ray spectrum.

For each of the seven field lines, we compute the magnetic field strength $B_0(s)$ and the height $z(s)$ above the photosphere as functions of position $s$ along the field line. These parameters serve as inputs for the Alfv\'{e}n wave turbulence model in Section~\ref{subsec: alfven wave turbulence model}.

\subsection{Alfv\'en Wave Turbulence Model} \label{subsec: alfven wave turbulence model}

We employ a 3D MHD model to simulate the generation, propagation, and dissipation of Alfv\'{e}n waves in an open magnetic flux tube that extends from the lower solar atmosphere to the outer corona. The model considers a single flux tube with a circular cross section, spanning from the base of the photosphere to a height of $3 \, R_{\odot}$. The flux tube radius $R\left(s\right)$ varies with position $s$ along its central axis, defined in the range $s_0 < s < s_0+L$, where $s_0$ is the position of the positive polarity footpoint and $L$ is the length of the field line. The open field is assumed to be very thin compared to its length, $R\left(s\right) \ll L$. The background magnetic field $\boldsymbol{B}_0 \left( x,y,s \right)$ depends on the Cartesian coordinates $x$, $y$, and the axial position $s$, where $x^2+y^2 < R^2 (s)$. Both the magnetic field strength $B_0\left(s\right)$ and plasma density $\rho_0 \left( s \right)$ are considered to be uniform over the open field cross section. At the photospheric footpoint, $s=s_0$, transverse motions cause random intermixing of field lines inside the flux tube. The typical small-scale, random footpoint motions are observed to have an rms velocity of about $\Delta v_\mathrm{rms} \sim 1.5$~km/s \citep{2012ApJ...752...48C}. Here, we choose our standard $\Delta v_\mathrm{rms}$ to be $1.48$~km/s. We also consider 0.74~km/s as the intermediate rms velocity. These two rms velocities are consistent with the observed nonthermal line broadening \citep{2014ApJ...786...28A}. The resulting footpoint motions generate Alfv\'{e}n waves that propagate along the flux tube and dissipate their energy in the chromosphere and corona, contributing to atmospheric heating. In addition to the two rms velocities above, we consider an extreme case with $\Delta v_\mathrm{rms} = 2.95$~km/s as a toy model to assess how elevated turbulence levels affect the resulting gamma-ray emission.

\begin{figure}[t]
    \centering
    \includegraphics[width=0.99 \columnwidth]{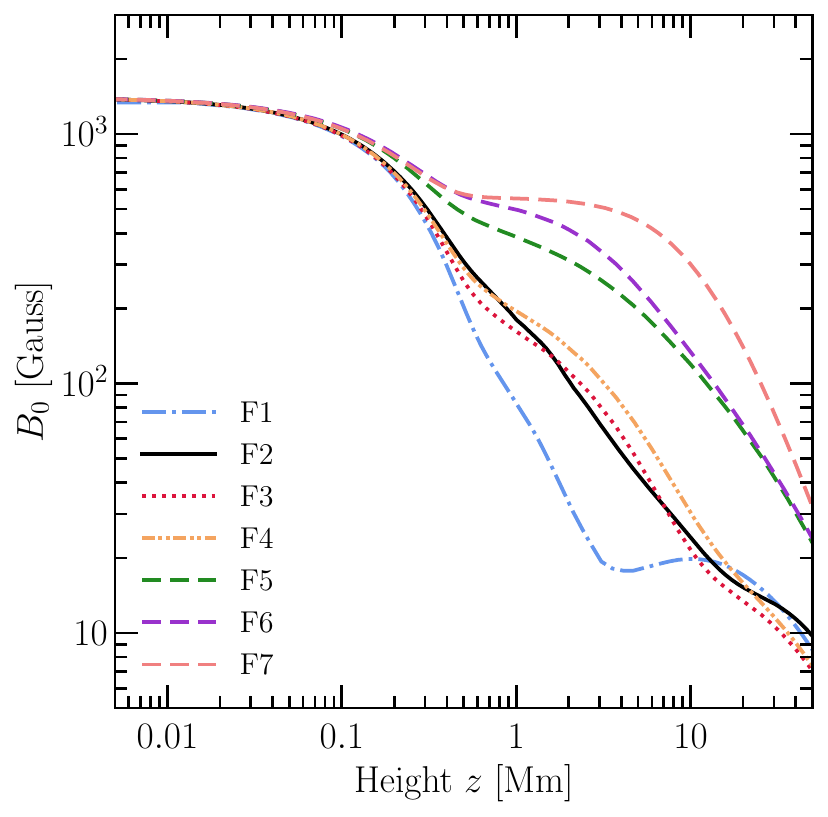}
    \caption{Background magnetic field strength $B_0$ for the seven magnetic network field lines calculated from the 3D MHD model.}
    \label{fig:all_lines_B0}
\end{figure}

\begin{figure*}[t]
    \centering
    \includegraphics[width=2.12 \columnwidth]{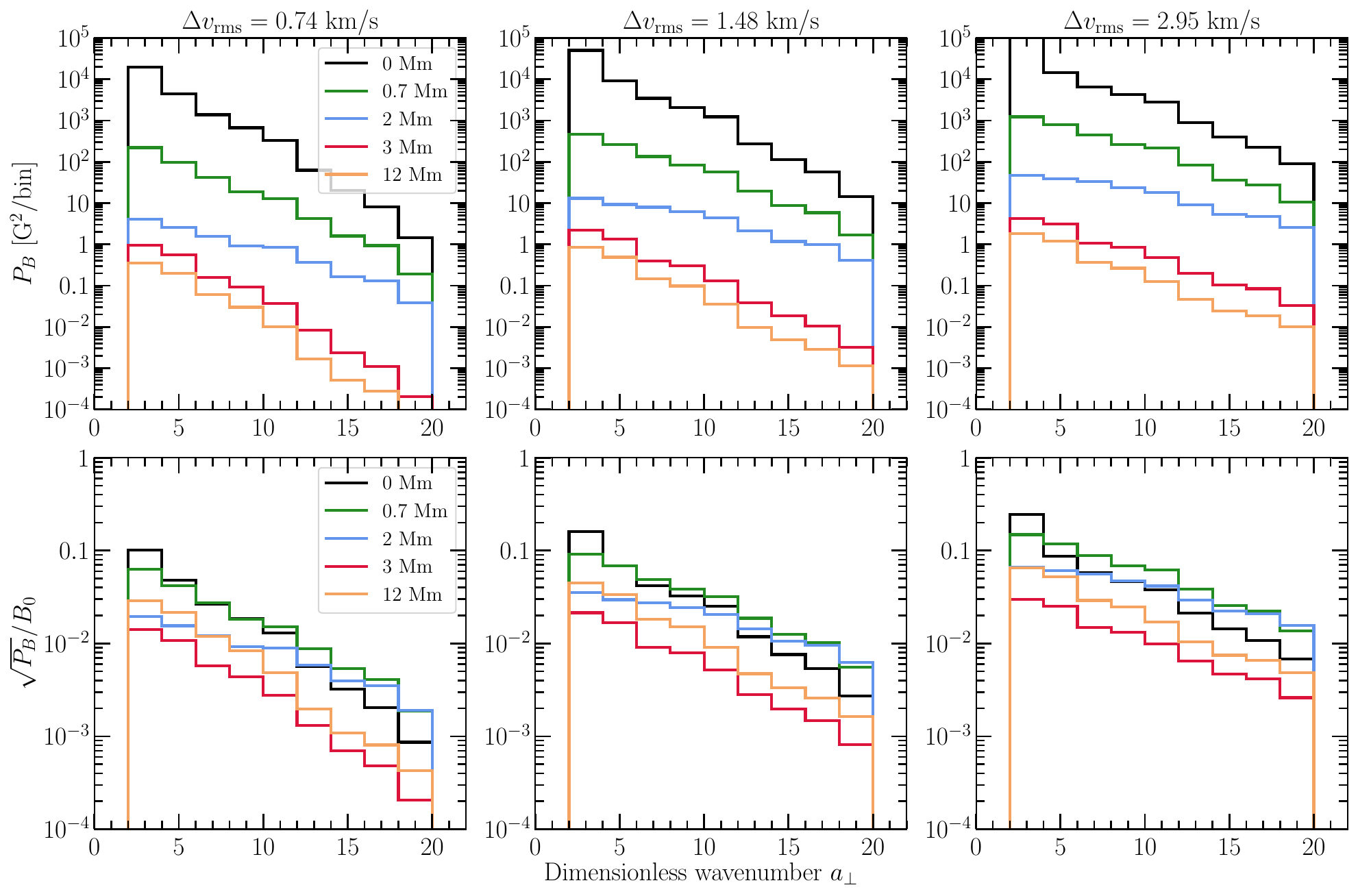}
    \caption{Magnetic power spectra (top row) and magnetic fluctuation amplitude (bottom row) within Network F2 case, shown for three $\Delta v_\mathrm{rms}$ conditions.}
    \label{fig:f2_PSD}
\end{figure*}

The 3D model adopts the RMHD approximation, which filters out slow- and fast-mode magnetoacoustic waves, retaining only Alfv\'{e}n waves \citep[e.g.,][]{1976PhFl...19..134S, 2011ApJ...736....3V}. Within this framework, magnetic perturbations $\boldsymbol{B}_1$ are assumed to be much smaller than the background field ($\lvert \boldsymbol{B}_1 \rvert \ll B_0$), and velocity fluctuations are considered small relative to the Alfv\'{e}n speed. Flows parallel to the background field are neglected. The magnetic perturbation is expressed as $\boldsymbol{B}_1 = \nabla_\perp h \times \boldsymbol{B}_0$, where $h\left( x,y,s,t \right)$ is the magnetic flux function. Similarly, the velocity field is given by $\boldsymbol{v} = \nabla_\perp f \times \boldsymbol{\hat{B}}_0$, where $f\left( x,y,s,t \right)$ is the velocity stream function, $\boldsymbol{\hat{B}}_0\left( x,y,s \right)$ is the unit vector along the background field, and $t$ is the time. Both $\boldsymbol{B}_1$ and $\boldsymbol{v}$ are perpendicular to $\boldsymbol{B}_0$, highlighting the highly anisotropic nature of the Alfv\'{e}n wave turbulence cascade. The evolution of the system is governed by two coupled equations for the stream function $f\left( x,y,s,t \right)$ and the flux function $h\left( x,y,s,t \right)$, given by:
\begin{equation}
\begin{aligned}
    \frac{\partial \omega} {\partial t} &+ \boldsymbol{\hat{B}}_0 \cdot
    \left( \nabla_\perp \omega \times \nabla_\perp f \right) \\
    &= v_A^2 \left[ \boldsymbol{\hat{B}}_0 \cdot \nabla \alpha +
    \boldsymbol{\hat{B}}_0 \cdot
    \left( \nabla_\perp \alpha \times \nabla_\perp h \right) \right] + D_v , 
    \label{eq:dodt}
\end{aligned}
\end{equation}
\begin{equation}
    \frac{\partial h} {\partial t} = \boldsymbol{\hat{B}}_0 \cdot \nabla f
    + \frac{f}{H_B} + \boldsymbol{\hat{B}}_0 \cdot \left( \nabla_\perp f \times
    \nabla_\perp h \right) + D_m ,
     \label{eq:dhdt}
\end{equation}
where $\omega$ ($\equiv - \nabla_\perp^2 f$) is the parallel vorticity, and $\alpha$ ($\equiv -\nabla_\perp^2 h$) is the magnetic twist parameter. The magnetic scale height is defined as $H_B\left(s\right) \equiv B_0/ \left(dB_0/ds\right)$, and the Alfv\'{e}n speed is given by $v_A\left(s\right) \equiv B_0 / \sqrt{4 \pi \rho_0}$. The terms $D_v$ and $D_m$ represent viscous and resistive damping, respectively. Following \citet{2011ApJ...736....3V}, the damping rates scale with the sixth power of the perpendicular wavenumber. The modes with the highest wavenumber have a damping rate $\nu_\mathrm{max} = 2.54~\mathrm{s^{-1}}$.

The nonlinear terms in the above equations involve derivatives with respect to the transverse coordinates~$x$ and~$y$, and are responsible for the transfer of wave energy to small spatial scales, a hallmark of turbulent cascades. This turbulent cascade of waves occurs in the presence of counter-propagating Alfv\'{e}n waves. \citet{2011ApJ...736....3V} demonstrated that strong wave reflection takes place in the chromosphere and the transition region. These reflections generate counter-propagating waves in the chromosphere, leading to strong turbulence in those regions of the flux structure. We simulated the development of such turbulence and found that most of the injected energy is dissipated in the chromosphere, consistent with observations. However, about $10\%$ of the wave energy is transmitted into the corona, where it contributes to the generation of coronal turbulence. 

In this study, we extend the Alfv\'{e}n wave turbulence model to seven open magnetic field lines, labeled F1 to F7 in Figure~\ref{fig:fieldline_visualization}, selected through the potential field extrapolation. For each field line, we  construct a 3D MHD model and simulate the dynamics of the Alfv\'{e}n waves within a thin tube surrounding the selected field line. Each flux tube is assumed to have a circular cross section. The magnetic field strength $B_0\left(s\right)$ is obtained from the potential field model, and used as the background field for the turbulence simulation. Because the potential field model does not capture the structure of the photospheric and chromospheric layers accurately, we modify the field strength profile $B_0\left(s\right)$ near the end of the open field. In particular, we impose a photospheric field strength $B_0\left(s_0\right) = 1400$~G at the footpoint, consistent with observational data. The radius $R\left(s\right)$ of the flux tube is set to vary in accordance with magnetic flux conservation, $B_0 R^2 =$ constant, and we assume $R\left(s_0\right) = 100$ km in the photosphere.

The dynamics of the Alfv\'{e}n waves within each flux tube is simulated by numerically solving the reduced MHD equations~\eqref{eq:dodt} and~\eqref{eq:dhdt}, following the methodology outlined in \citet{2011ApJ...736....3V}. The waves are generated by random footpoint motions with rms velocities of 0.74, 1.48, and 2.95~km/s \citep{2011ApJ...743..133A, 2012ApJ...752...48C, 2014ApJ...786...28A} and correlation time $\tau_c = 60$~s. The spatial grid resolves transverse structure to approximately one-tenth of the flux tube radius, allowing for accurate modeling of perpendicular wave dynamics. Simulations have a temporal resolution of about 0.1 s, over a total duration of 3000 s, significantly longer than the Alfv\'{e}n travel time along the full length of each field line. For each flux tube, the modeling provides detailed information about the dynamics of the Alfv\'{e}n waves and the turbulent heating produced by these waves.

Plasma temperature $T_0\left(s\right)$ varies rapidly with position in the transition region at the end of the field line. According to solar atmospheric models \citep[e.g.,][]{1999ApJ...518..480F}, the temperature in the lower transition region varies on a length scale of only a few kilometers. The Alfv\'{e}n speed in the lower transition region is quite high ($\sim 200$~km/s) and varies rapidly with position along the field line. In our simulations, Alfv\'{e}n waves propagating upward from the chromosphere strongly reflect at the transition region; numerical aspects of such reflection are described in Appendix~B of \citet{2011ApJ...736....3V}.

To perform the 3D RMHD modeling, we begin by constructing a 1D background atmosphere inside the open field. The magnetic field configuration is an input of the model, derived using the method described in the previous section. We also choose an initial coronal pressure  $p_\mathrm{cor}$, and estimate the maximum temperature  $T_\mathrm{max}$, based on the Rosner-Tucker-Vaiana scaling law \citep{1978ApJ...220..643R}. These values are used to construct a background atmosphere. Then, the Alfv\'{e}n waves are simulated over a duration of 3000~seconds.

Figures~\ref{fig:all_lines_B0} and~\ref{fig:f2_PSD} present the Alfv\'{e}n wave turbulence modeling results. Figure~\ref{fig:all_lines_B0} shows the background magnetic field $B_0$ as a function of height $z$ for the Network F1 to F7. Figure~\ref{fig:f2_PSD} shows the magnetic power spectra (top row), $P_B\left(z, a_\perp\right)$, and the magnetic fluctuation amplitude (bottom row), $\sqrt{P_B\left(z, a_\perp\right)} / B_0\left(z\right)$, as a function of dimensionless perpendicular wavenumber $a_{\perp} \equiv k_\perp  R\left( z \right)$, where $k_\perp$ is the dimensional perpendicular wavenumber. Here, we show the Network F2 case at five heights in each of the three footpoint rms velocities $\Delta v_\mathrm{rms} = 0.74$, $1.48$, and $2.95~\mathrm{km/s}$. The perpendicular wavenumber is derived directly from the RMHD formalism along selected open magnetic field lines (e.g., Network F2). The wavenumber $a_{\perp}$ is derived from eigenmodes of a cylindrical flux tube and is used to compute magnetic power spectra at various heights and turbulence strengths \citep{2011ApJ...736....3V}.

Last, we note that the field-aligned coordinate $s$ does not generally coincide with the vertical height $z$, as the field line geometry is determined by the ambient magnetic field distribution (see Figure~\ref{fig:fieldline_visualization}). As the RMHD background atmosphere is evaluated along the field-aligned coordinate $s$, solar atmospheric quantities such as $\rho_0(s)$ and $T(s)$ are different from the purely height-dependent profiles, $\rho_0(z)$ and $T(z)$, typically found in 1D solar models.


\section{Magnetostatic Model} \label{sec: magnetostatic model}

In this section, we present the basic equations and numerical solutions for the magnetostatic flux structures considered in this study, following the methodology of \citet{1986A&A...170..126S} and Paper~\citetalias{2024ApJ...961..167L}. We develop and present two main classes of solutions: (1) a tube-like solution that models the overall network field and individual flux tubes, and (2) a sheet-like solution for the intergranular sheet. The detailed numerical results for these configurations are illustrated using the Network F2 case in Figure~\ref{fig:fieldline_visualization}; the other four cases are not shown.

We note that the symbol $\boldsymbol{B}_{0}$ used in this section refers to the magnetic field calculated from the magnetostatic method. It should be distinguished from the background magnetic field notation used in the RMHD method in Section~\ref{sec: Alfven wave model}.

\subsection{Equations for Magnetostatic Flux Structures}

First, we consider a magnetic tube-like structure designed to model the network field and flux tube. The structure is assumed to be vertical, untwisted, and azimuthally symmetric. This structure is embedded in a gravitationally stratified, field-free medium and is maintained in magnetostatic equilibrium with the surrounding gas. We solve the structure in the cylindrical $r\phi z$-coordinate system where $r$ is the radial distance, $\phi$ is the azimuth angle, and $z$ is the axial height. The azimuthal symmetry (i.e., $\partial \boldsymbol{B}_{0} / \partial \phi = 0$) implies that the vector potential $\boldsymbol{A}$ possesses only an $A_\phi$ component. The magnetic field $\boldsymbol{B}_{0} \left(r, z\right)$ can then be conveniently described using a stream function, $\Psi \left(r, z\right) \equiv r A_\phi$, yielding the components:
\begin{equation}
    B_{0, r}\left(r, z\right) = -\frac{1}{r} \frac{\partial \Psi}{\partial z}, \quad B_{0, z}\left(r, z\right) = \frac{1}{r} \frac{\partial \Psi}{\partial r}.
\end{equation}
Expressed in terms of $\Psi$, Amp{\`e}re's Law takes the quasi-linear form (Grad-Shafranov):
\begin{equation}
    \frac{\partial^2 \Psi}{\partial r^2} - \frac{1}{r} \frac{\partial \Psi}{\partial r} + \frac{\partial^2 \Psi}{\partial z^2} = -4\pi J_\phi
    \label{eq: Grad-Shafranov tube-like}
\end{equation}
where $J_\phi$ is the azimuthal component of the volume current density $\boldsymbol{J}$ (with the dimension of [$\mathrm{current/length^2}$]). Since $J_\phi$ is generally a function of $\Psi$, Equation~\eqref{eq: Grad-Shafranov tube-like} is nonlinear and typically solved iteratively subject to specified boundary conditions. The solution $\Psi \left(r, z\right)$ of Equation~\eqref{eq: Grad-Shafranov tube-like} is non-zero within the tube-like structure of radius $R\left(z\right)$ and zero outside this boundary.

Next, we consider a sheet-like magnetic structure to represent an intergranular sheet. This structure is also assumed to be vertical and untwisted, and it is maintained in magnetostatic equilibrium with the ambient gas. Its magnetic field is assumed to be translationally invariant in one horizontal direction (e.g., along the $x$-axis in a Cartesian $xyz$-coordinate system). In this case, $\Psi \left(y, z\right) \equiv A_y$, and the $\boldsymbol{B}_{0}$ components are given by
\begin{equation}
    B_{0, y}\left(y, z\right) = -\frac{\partial \Psi}{\partial z}, \quad B_{0, z}\left(y, z\right) = -\frac{\partial \Psi}{\partial y}.
\end{equation} 
Amp{\`e}re's Law can then be written in quasi-linear form as
\begin{equation}
    \frac{\partial^2 \Psi}{\partial y^2} + \frac{\partial^2 \Psi}{\partial z^2} = - 4\pi J_x,
    \label{eq: Grad-Shafranov sheet-like}
\end{equation}
where $J_x$, the $x$-component of $\boldsymbol{J}$, is also a function of $\Psi$. Consequently, this equation is also nonlinear and is generally solved using iterative methods constrained by appropriate boundary conditions. The solution $\Psi \left(y, z\right)$ of Equation~\eqref{eq: Grad-Shafranov sheet-like} is non-zero within the sheet-like structure of half-width $Y\left(z\right)$ (for $y \geq 0$) and zero outside this boundary.

To solve the quasi-linear Equations~\eqref{eq: Grad-Shafranov tube-like} and \eqref{eq: Grad-Shafranov sheet-like}, we must first determine $\boldsymbol{J}$. This, in turn, requires a description of the gas temperature $T$ and pressure $P$. We assume a simplified thermal profile where the temperature $T\left(z\right)$ is horizontally uniform at any given height $z$, i.e., $T_i\left(z\right) = T_e\left(z\right) = T\left(z\right)$; subscripts $i$ and $e$ denote the interior and exterior of the flux structure, respectively. Consequently, the pressure scale height $H_P$ is the same for both interior and exterior regions and is expressed as
\begin{equation}
    H_P\left(z\right) = \frac{R_\odot^2 k_b T}{G M_\odot \overline{m}},
\end{equation}
where $R_\odot$ is the solar radius, $k_b$ is the Boltzmann constant, $G$ is the Newton constant, $M_\odot$ is the solar mass, and $\overline{m}$ is the mean particle mass of the solar gas. As a result, the interior and exterior gas pressures are given as
\begin{equation}
    P_i\left(z\right) = P_i\left(z_b\right) \exp \bigg[-\int_{z_b}^z \frac{dz^\prime}{H_P\left(z^\prime\right)}\bigg],
    \label{eq: P_i}
\end{equation}
\begin{equation}
    P_e\left(z\right) = P_e\left(z_b\right) \exp \bigg[-\int_{z_b}^z \frac{dz^\prime}{H_P\left(z^\prime\right)}\bigg],
    \label{eq: P_e}
\end{equation}
where $z_b$ denotes the base of the computation box. At the boundary surface of the flux structure, the continuity of total pressure, $P + B_{0}^2 / 8 \pi$, requires
\begin{equation}
    P_i + B_{0, i}^2 / 8 \pi = P_e + B_{0, e}^2 / 8 \pi.
    \label{eq: total pressure continuity}
\end{equation}

Given that the pressure inside the flux structure depends solely on height $z$, the internal gas satisfies hydrostatic equilibrium, $\nabla P - \rho \boldsymbol{g} = 0$, resulting in $\boldsymbol{J} = 0$ everywhere inside the flux structure. However, a discontinuity in the magnetic field at the boundary surface generates a surface current density, $J^\star$ (with the dimension of [current/length]). This is related to a jump in magnetic fields by $J^\star = \left(B_{0, i} - B_{0, e}\right)/4\pi$, and by applying the total pressure continuity in Equation~\eqref{eq: total pressure continuity}, it can also be expressed as $J^\star = 2 \left(P_e - P_i\right) / \left(B_{0, e} + B_{0, i}\right)$.

\subsection{Boundary Conditions and Atmospheric Models} \label{subsec: boundary conditions}

The quasi-linear Equations~\eqref{eq: Grad-Shafranov tube-like} and~\eqref{eq: Grad-Shafranov sheet-like} are solved iteratively using the finite-difference method detailed in \citet{1986A&A...170..126S}. Numerical stability during this iterative process is achieved by applying the implicit relaxation method described in \citet{1980wdch.book.....P}. Due to symmetry, for the tube-like structure, we only need to solve in the $rz$-plane. For the sheet-like structure, we similarly only solve in the $yz$-plane for the region where $y \geq 0$. In both cases, the resulting 2D magnetic field is then projected onto the 3D $xyz$ coordinate system. The resulting 3D magnetic field is then used for the GCR trajectory calculations presented in Section~\ref{sec: simulation setup}.

For each of the seven network field line cases investigated, the network field from $700$~km to $12$~Mm, and the flux tube extends from $-200$~km to $700$~km. 

A single model for the intergranular sheet, which ranges in height from $-200$~km to $600$~km, is utilized for all seven field line cases. The motivation for setting the upper boundary of the intergranular sheet at $600$~km is explained in the following paragraph.

Our model excludes the subcanopy domain, which is the region between the top of the intergranular sheet ($z=600$~km) and the canopy field ($z \sim 2-4$~Mm). This limitation is due to the fact that accurately modeling the magnetic structure in this domain requires 2D or 3D magnetoconvection simulations \citep[e.g.,][]{2005A&A...429..335V, 2005ESASP.596E..65S, 2006ASPC..354..345S, 2009A&A...504..595Y}, which are beyond the scope of the magnetostatic Grad-Shafranov equations used in this work. However, our choice of choosing the upper boundary of the sheet at $600$~km can be justified by two key findings from previous magnetoconvection studies. First, \citet{2009A&A...504..595Y} shows that the sheet-like structures in their MURaM 3D magnetoconvection simulation below the height of 600~km can be well described by the magnetostatic approach. Moreover, \citet{2006ASPC..354..345S} shows that the magnetic field strength in the subcanopy domain is mostly below 40~G, which is much smaller than the kilo-Gauss sheet structures at the photospheric layer. These findings support our method of applying the height of the intergranular sheet up to $z=600$~km in our magnetostatic Grad-Shafranov equations.

Solving the quasi-linear Equations~\eqref{eq: Grad-Shafranov tube-like} and~\eqref{eq: Grad-Shafranov sheet-like} requires the prior specification of boundary conditions. For all modeled structures (i.e., network field, flux tube, and intergranular sheet), the magnetic field at the top of the computational domain is assumed to be purely vertical, resulting in a Neumann boundary condition, $\partial \Psi / \partial z = 0$. At the base of the computational domain ($z=z_b$), the vertical magnetic field component $B_{0z}\left(z_b\right) \equiv B_{0z}^{\star}$ is uniform inside the boundary surface (which for tube-like structure is located at $r = R_{\star}$ and for sheet-like structure at $y = Y_\star$), while the magnetic field outside the boundary surface is set to zero. Additionally, the magnitude of $B_{0z}^{\star}$ also serves as a boundary condition and is treated as an input parameter when solving the quasi-linear Equations~\eqref{eq: Grad-Shafranov tube-like} and~\eqref{eq: Grad-Shafranov sheet-like}. We elaborate on our choice for $B_{0z}^{\star}$ in Section~\ref{subsec: numerical magnetostatic}.

\begin{figure}[t]
    \centering
    \includegraphics[width=0.99 \columnwidth]{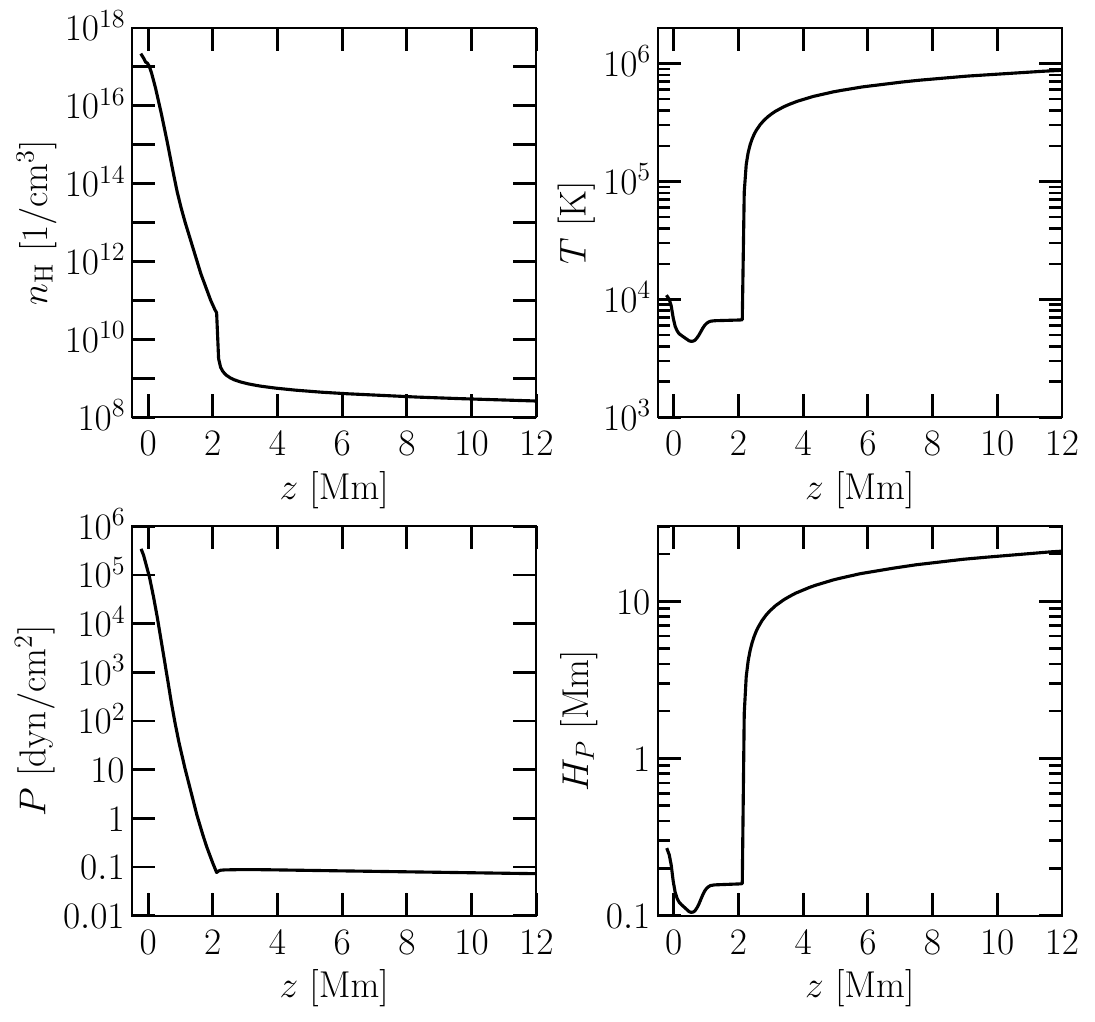}
    \caption{Solar atmosphere profiles used in this work.}
    \label{fig:atmosphere_profile}
\end{figure}

Figure~\ref{fig:atmosphere_profile} shows the background atmospheric profiles for $P$, $T$, $H_P$, and $n_\mathrm{H}$. The data above $z=-100$~km follow the average atmospheric quiet-Sun model of \citet{2008ApJS..175..229A}. The data between $z=-200$~km and $z=-100$~km follow the HRSASP model from \cite{1979ApJ...232..923C}. The $P$, $T$, and $H_P$ profiles are the input for solving the quasi-linear Equations~\eqref{eq: Grad-Shafranov tube-like} and~\eqref{eq: Grad-Shafranov sheet-like}, while the $n_\mathrm{H}$ profile will be used to calculate the gamma-ray emission in Section~\ref{sec: simulation setup}. Finally, for all height ranges we considered, we set the mean particle mass $\overline{m}$ to $1.273~\mathrm{a.m.u.}$. This value is based on the work of \citet{2008ApJS..175..229A} and is consistent with the findings from \citet{1979ApJ...232..923C}.

\begin{figure}[t]
    \centering
    \includegraphics[width=0.99 \columnwidth]{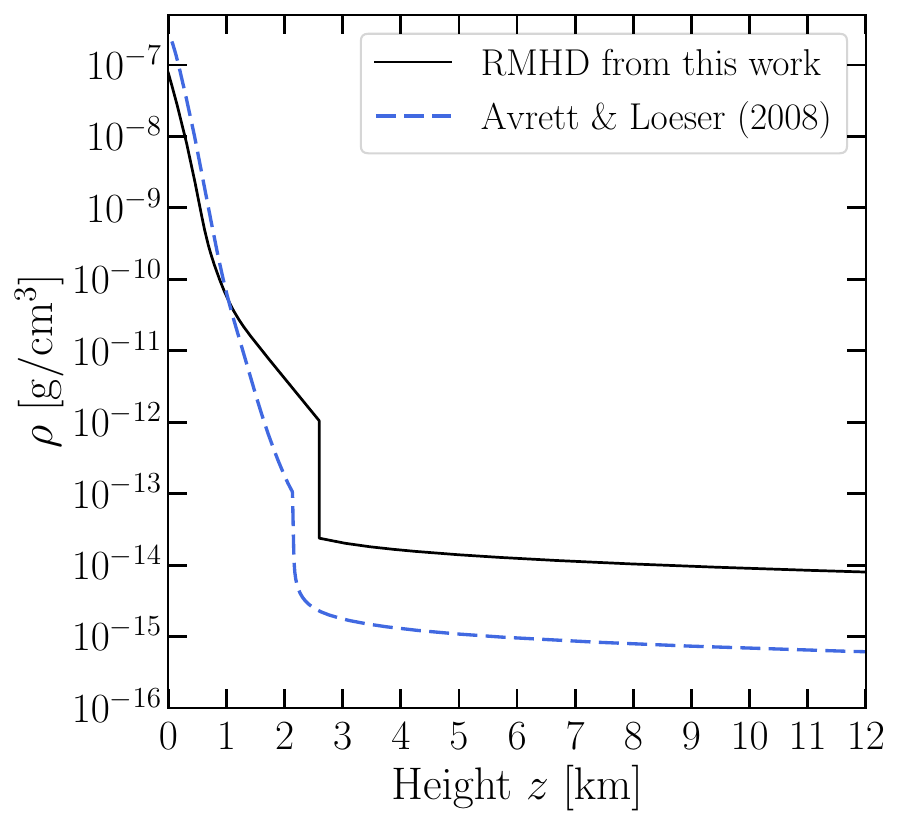}
    \caption{Comparison of the mass density for the RMHD F2 Field Line case and the semi-empirical quiet-Sun density profile of \citet{2008ApJS..175..229A}.}
    \label{fig:rho_comparison}
\end{figure}

Figure~\ref{fig:rho_comparison} compares the mass density profile as a function of height $z$ for the RMHD Network F2 simulation case with the semi-empirical quiet-Sun model of \citet{2008ApJS..175..229A}. Unlike the plane-parallel, static atmosphere assumed by \citet{2008ApJS..175..229A}, the RMHD background atmosphere is constructed to satisfy force balance and Alfv\'{e}n wave propagation within an expanding magnetic flux tube subject to coronal boundary conditions. Consequently, some divergence between the two models is expected. As demonstrated later in this work, gamma-ray emission primarily occurs in the region between $z=-100$~km and $400$~km; within this range, the density profile from RMHD method is consistent with the semi-empirical model of \citet{2008ApJS..175..229A}.

\subsection{Numerical Solutions} \label{subsec: numerical magnetostatic}

We first present the parameters and numerical results for the network field and flux tube associated with F2. While the same methodology was applied to four other field line cases, their specific results are not presented here for brevity. Subsequently, we present the results for the intergranular sheet, which is analyzed independently of all seven field line cases.

\subsubsection{Network Field Line~2}

In the context of the Network F2 case, $B_{0z}^{\star}$ and $R_{\star}$ are the values of the vertical magnetic field component $B_{0z}$ and the network radius $R$, respectively, evaluated at the network field lower boundary at the height $z_b=700$~km. The numerical values of $B_{0z}^{\star}$ and $R_{\star}$ were chosen to ensure the horizontally-averaged magnetic field, $\bar{B}\left(z\right)$, from the magnetostatic method closely matches the background magnetic field $B_0$ from the RMHD method. At $z_b=700$~km, the input $B_{0z}^{\star}$ for the magnetostatic calculation is equated to the $B_0$ value from the RMHD simulation in Figure~\ref{fig:all_lines_B0}. Applying this procedure to the Network F2 case yields $B_{0z}^{\star} = 13.6$~G at $z=700$~km.

Once $B_{0z}^{\star}$ fixed, the parameter $R_{\star}$ is linked to the total magnetic flux, $\Phi_B$, through the relation $\Phi_B = \pi R_{\star}^2 B_{0z}^{\star}$. For clarity and convenience in the subsequent discussion, $\Phi_B$ is adopted as the primary fitting parameter instead of $R_{\star}$. To determine the optimal $\Phi_B$, we perform numerical calculations using the quasi-linear Equation \eqref{eq: Grad-Shafranov tube-like} for a range of $\Phi_B$ values, sampled at intervals of $10^{19}$~Mx. The value of $\Phi_B$ that yields a $\bar{B}\left(z=12~\text{Mm}\right)$ most consistent with the RMHD-derived solution $B_0\left(z=12~\text{Mm}\right)$ is then selected. In the Network F2 case, we find that $\Phi_B = 3 \times 10^{19}$~Mx, corresponding to $R_{\star} = 2$~Mm at $z=700$~km, provides the best agreement.

\begin{figure}[t]
    \centering
    \includegraphics[width=0.99 \columnwidth]{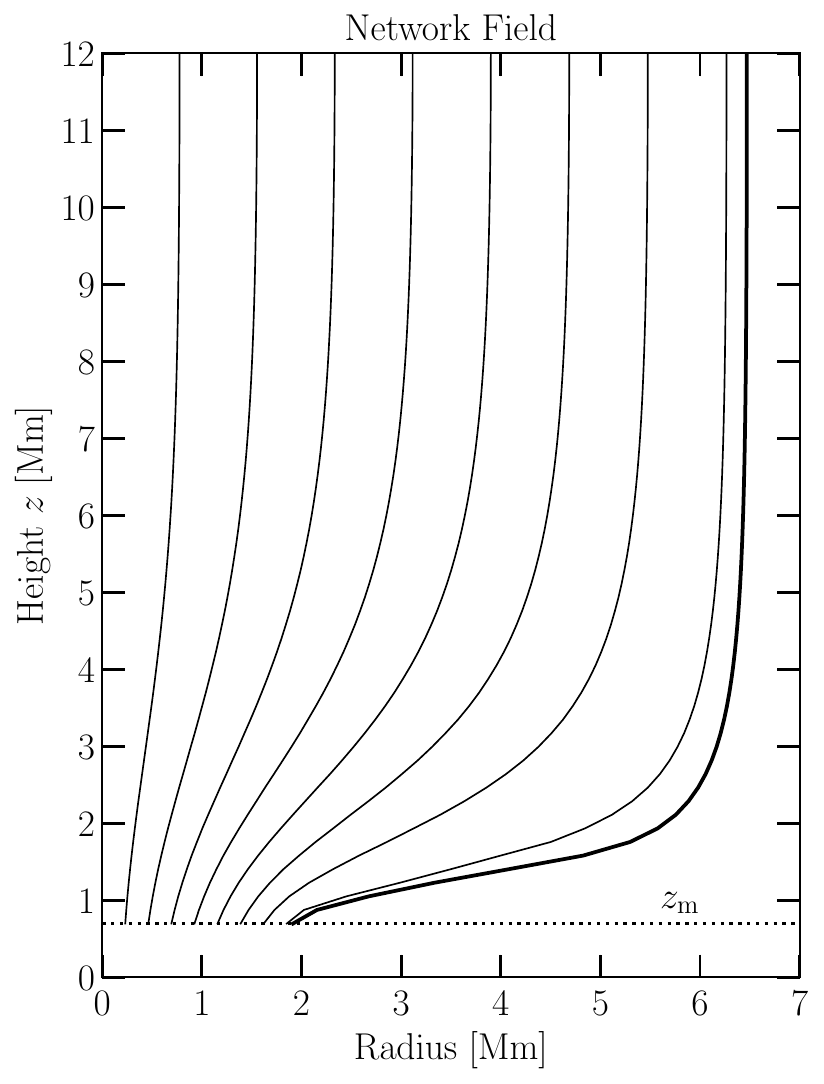}
    \caption{Magnetostatic network field structure for the Network F2 case. Horizontal dotted line ($z_\mathrm{m} = 0.7$~Mm) is the merging height of flux tubes, which also marks the lower boundary of the network field. Thin lines, progressing from smaller to larger radii, denote the magnetic field lines enclosing magnetic flux valued at $1^2, 2^2, 3^2, \ldots$ times a single flux tube's magnetic flux. Thick line is the outer edge of the network field.}
    \label{fig:magnetostatic_NF}
\end{figure}

Figure~\ref{fig:magnetostatic_NF} shows the resulting magnetostatic structure of the network field for the Network F2 case, with the input parameters $B_{0z}^{\star} = 13.6$~G and $R_{\star} = 2$~Mm at $z_b = 700$~km. The computation domain is a $11.3~\mathrm{Mm} \times 7~\mathrm{Mm}$ box discretized on a rectangular mesh of $65 \times 65$ mesh points. Solid thin lines represent the magnetic field lines composing the network structure. From $z=0.7$~Mm to $z \approx 2$~Mm, the network structure expands laterally as height increases. Within this height range, the predominantly horizontal magnetic field at the structure's periphery (solid thick line) forms the canopy field. Subsequently, from $z \approx 2$~Mm to $z \approx 4$~Mm, this network structure begins to merge with adjacent network structures. Above $z \approx 4$~Mm, these individual network structures fully coalesce, and their combined magnetic field occupies the entire available volume, expanding primarily in the vertical direction.

For the flux tube in the Network F2 case, the parameters $B_{0z}^{\star}$ and $R_{\star}$ represent the vertical magnetic field component $B_{0z}$ and the flux tube radius $R$, respectively, evaluated at the height $z_b=-200$~km. The numerical values of these input parameters are chosen so that the resulting $\bar{B}$ and $R$ at the photosphere base at $z=0$~km are approximately $1400$~G and $100$~km, respectively. These target values are selected to be consistent with the results from our RMHD method in Section~\ref{sec: Alfven wave model}. In order to determine the input $B_{0z}^{\star}$ value, we use the thin-tube approximation, which states that the magnetic field strength in the flux tube varies with height as $e^{-z/2H_p\left(z\right)}$~\citep{2005ApJS..156..265C}. Integrating from $z=0$~km to $z=-200$~km, we have $B_{0z}^{\star} \approx 2255$~G. Subsequently, using flux conservation, the flux radius at this depth is found to be $R_{\star} \approx 79$~km. Last, we note that matching at $z = 0$~km with $\bar{B} = 1400$~G and $R = 100$~km yields a single solution for the flux tube. Therefore, the uncertainty in fitting the magnetostatic solution to the RMHD-derived mean field for the flux tube case cannot be quantified.

\begin{figure}[t]
    \centering
    \includegraphics[width=0.99 \columnwidth]{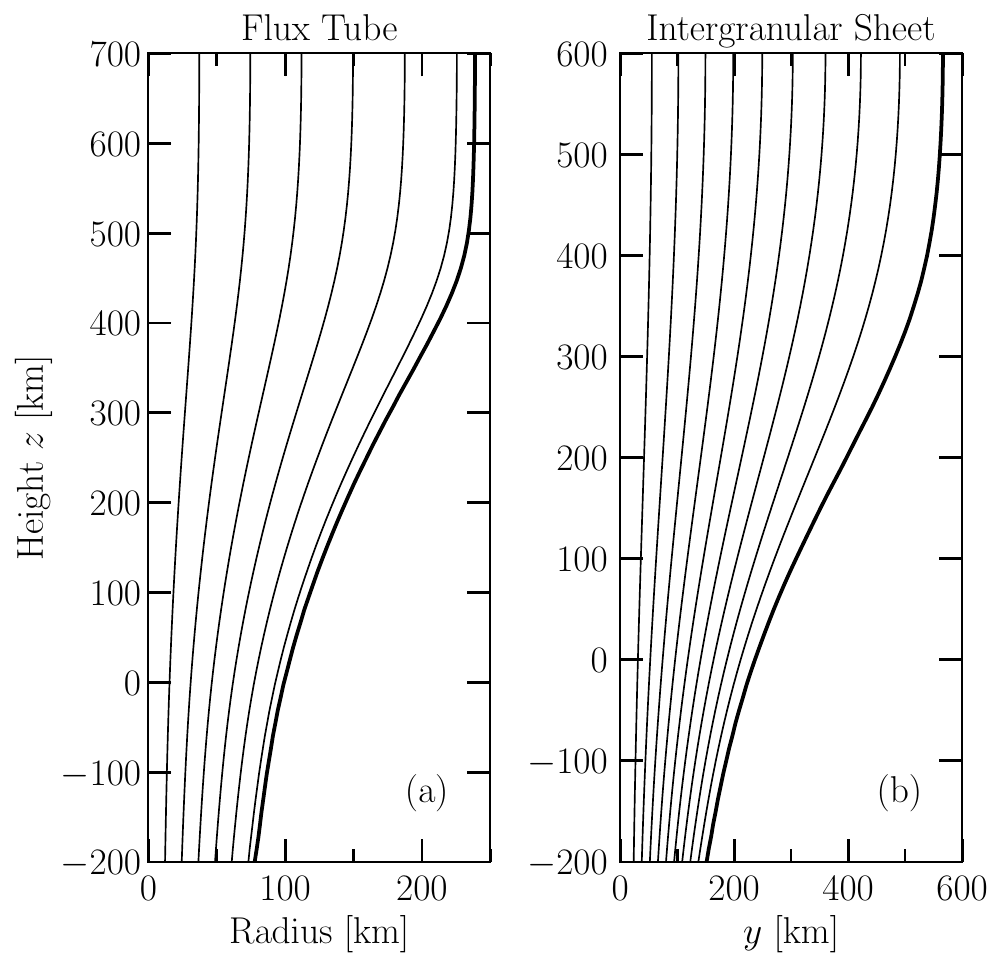}
    \caption{(a) Magnetostatic model of the flux tube structure for Network F2. (b) Magnetostatic model of the intergranular sheet structure for all seven field line cases.}
    \label{fig:magnetostatic_NE_IGS}
\end{figure}

Figure~\ref{fig:magnetostatic_NE_IGS}(a) shows the resulting magnetostatic structure of the individual flux tube for the Network F2 case. The input parameters are $B_{0z}^{\star} = 2255$~G and $R_{\star} = 80$~km. The computation domain is a $\mathrm{900~km} \times \mathrm{247.5~km}$ box discretized on a rectangular mesh of $65 \times 65$ mesh points. In the uppermost convection zone and the photosphere base, the magnetic field is concentrated in only a small fraction of space. As gas pressure decreases with increasing height, individual flux tubes tend to expand and merge with nearby counterparts at $\mathrm{500~km} \lesssim z \lesssim \mathrm{700~km}$. This merging process leads to the formation of a larger, monolithic network field structure. For the Network F2 case, $N_\mathrm{tube} = 66$ such flux tubes coalesce to form the network field structure of $\Phi_B = 3 \times 10^{19}$~Mx shown in Figure~\ref{fig:magnetostatic_NF}. The characteristics of all seven network field lines are shown in Table~\ref{table: field_line_characteristics}.

In the context of this merging flux tube framework, each flux tube can be treated as an equivalent magnetic field line. This definition aligns with our usage of ``field line'' in the RMHD method in Section~\ref{sec: Alfven wave model}. Consequently, the network field structure is conceptualized as a collection of $N_\mathrm{tube}$ identical field lines.

However, in the RMHD method, the background magnetic field strength $B_0\left(s\right)$ is considered to be uniform over the flux structure cross section and does not resolve the transverse (2D) structure. We should therefore interpret $B_0\left(s\right)$ as the cross-sectional mean magnetic field at axial position $s$. To recover the internal transverse structure of a network field line, we employ the magnetostatic model introduced in this section. The Network F2 case is illustrated in Figures~\ref{fig:magnetostatic_NF} and~\ref{fig:magnetostatic_NE_IGS}(a).

\begin{figure}[t]
    \centering
    \includegraphics[width=0.99 \columnwidth]{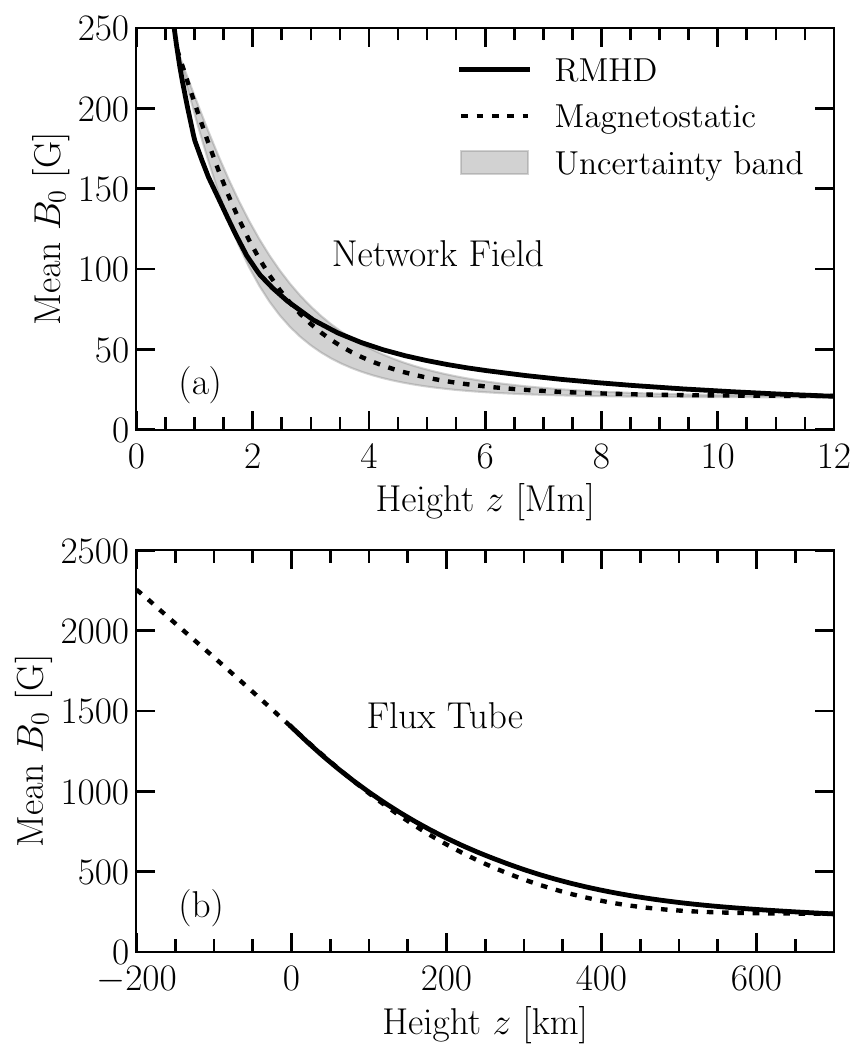}
    \caption{Comparison of $B\left(r\right)$ from RMHD and magnetostatic solutions for the Network F2 case: (a) in the network field, and (b) in the flux tube.}
    \label{fig:f2_B0_NF_NE_comparison}
\end{figure}

Figure~\ref{fig:f2_B0_NF_NE_comparison} shows a comparison between $B_0\left(z\right)$ obtained from the RMHD method and $\bar{B}\left(z\right)$ from the magnetostatic method, for both network field and flux tube structures in the Network F2 case. In panel~(a) for the network field, the gray band denotes the range of $\bar{B}\left(z\right)$ obtained by varying $\Phi_B$ by $\pm 10^{19}$~Mx about the fiducial $3\times 10^{19}$~Mx. In panel~(b) for the flux tube, a fitting uncertainty is not defined in our framework because enforcing $\bar{B} = 1400$~G and $R = 100$~km at $z = 0$~km yields a single solution. From $z=0$~km to $z=12$~Mm, the two approaches agree closely. This consistency validates our combined approach, which utilizes the RMHD method for Alfv\'{e}n wave turbulence calculations and the magnetostatic method to construct the 3D magnetic flux structures and their mean fields.

\subsubsection{Intergranular magnetic sheet}

For the intergranular magnetic sheet, the parameters $B_{0z}^{\star}$ and $Y_{\star}$ represent the vertical magnetic field component $B_{0z}$ and the sheet's half-width $Y$, respectively, evaluated at the lower boundary of the intergranular sheet of the height $z_b=-200$~km. The numerical values of these input parameters are chosen so that the resulting $\bar{B}$ and $Y$ at the photosphere base at $z=0$~km are approximately $1400$~G and $200$~km, respectively. These target values are selected to be consistent with the broad flux sheet concentration result from \citet[][Section~5.2]{2009A&A...504..595Y}, which utilized the \texttt{MURaM} 3D magneto-convection simulation code \citep{2005A&A...429..335V} to analyze the structure of small vertical photospheric magnetic flux concentrations. Similar to the method for the flux tube, we also use the thin-tube approximation and flux conservation, finding that the desired input parameters are $B_{0z}^{\star} = 2255$~G and $Y_{\star} = 150$~km.

The upper boundary of the computational domain for the intergranular sheet is set at $z=600$~km. This limitation is imposed because the quasi-linear approach in Equation~\eqref{eq: Grad-Shafranov sheet-like} requires a Neumann boundary condition that enforces a vertical magnetic field at this upper boundary. Observations from 2D and 3D magnetoconvection simulations support this, showing predominantly vertical magnetic fields in intergranular lanes at the photosphere. However, these simulations also reveal the presence of small-scale, horizontal magnetic canopy fields ($B \lesssim 100$~G) at $z\sim 600$~km. These canopy fields connect regions of opposite polarity in adjacent intergranular lanes, spanning distances comparable to a granular scale. Beneath these canopies and between intergranular lanes, magnetic voids with significantly weaker fields $B \lesssim 3$~G have been found \citep{2005ESASP.596E..65S, 2006ASPC..354..345S}.

As we show in our numerical results in Section~\ref{sec: results}, gamma rays escaping the solar surface originate primarily within the height range $-100~\mathrm{km} \lesssim z \lesssim 400~\mathrm{km}$. Within this region, strong magnetic fields are concentrated in the vertical structures, distinct from the aforementioned magnetic voids. Furthermore, GCRs that penetrate the network field to reach intergranular sheets possess high energies ($E_p^\mathrm{k} \gtrsim 1$~TeV) and are thus primarily influenced by strong magnetic fields. These considerations justify our simplified model of a vertical intergranular sheet extending from $z = -200$~km to $z = 600$~km. A comprehensive model of the complete internetwork magnetic field structure, including canopy loops, would require full 3D magnetoconvection simulations (e.g., with the \texttt{MURaM} code), as further discussed in Section~\ref{sec: results}.

Figure~\ref{fig:magnetostatic_NE_IGS}(b) shows the resulting magnetostatic structure of the intergranular magnetic sheet in the $yz$ cross section. (The structure is redundant in the $x$ direction.) The computation domain is a $\mathrm{800~km} \times \mathrm{605~km}$ box discretized on a rectangular mesh of $65 \times 65$ mesh points. Similar to flux tubes, the magnetic field in the uppermost convection zone and photosphere base is primarily vertical and also concentrated in a small fraction of space. 

Finally, for the network field, flux tube, and flux sheet structures shown in Figures~\ref{fig:magnetostatic_NF} and~\ref{fig:magnetostatic_NE_IGS}, we compute the 2D magnetic field between mesh points using 2D linear interpolation. We then project this field into the 3D $xyz$ coordinate system, yielding a magnetostatic solution that is continuous in space rather than defined only at mesh points.

\begin{deluxetable}{cccccccc}
\label{table: field_line_characteristics}
\tablecolumns{4}
\tablewidth{1\columnwidth}
\tablecaption{Characteristics of the seven network field lines. $N_\mathrm{tube}$ denotes the number of flux tubes in a network element. $R_\star^\mathrm{tube}$, $R_\mathrm{surface}^\mathrm{tube}$, and $R_\mathrm{top}^\mathrm{tube}$ denote the radius of the flux tube at $z=-200$~km, 0~km, and $700$~km, respectively. $R_\star^\mathrm{NW}$ and $R_\mathrm{top}^\mathrm{NW}$ denote the radius of the magnetostatic network field at $z=700$~km and 12~Mm, respectively.}
\tablehead{
\colhead{Network} &
\colhead{$\Phi_B$~[Mx]} &
\colhead{$N_\mathrm{tube}$} &
\colhead{$R_\star^\mathrm{tube}$~[km]} &
\colhead{$R_\mathrm{surface}^\mathrm{tube}$~[km]} & 
\colhead{$R_\mathrm{top}^\mathrm{tube}$~[km]} & 
\colhead{$R_\star^\mathrm{NW}$}~[km] &
\colhead{$R_\mathrm{top}^\mathrm{NW}$~[km]} 
}
\startdata
F1 & $5 \times 10^{18}$ & 11   & 80 & 99  & 330 & 1133 & 2803     \\
F2 & $3 \times 10^{19}$ & 66   & 80 & 99  & 238 & 2009 & 6470     \\
F3 & $5 \times 10^{19}$ & 110  & 80 & 98  & 258 & 2853 & 9154     \\
F4 & $8 \times 10^{19}$ & 176  & 80 & 100 & 245 & 3325 & 9948     \\
F5 & $3 \times 10^{20}$ & 662  & 80 & 99  & 178 & 4728 & 9532     \\
F6 & $4 \times 10^{20}$ & 882  & 80 & 102 & 209 & 6193 & 11419    \\
F7 & $6 \times 10^{20}$ & 1323 & 80 & 101 & 178 & 5858 & 8662     \\
\enddata
\end{deluxetable}

\section{Simulation Setup for GCR Propagation and Gamma-Ray Emission} \label{sec: simulation setup}

In this section, we detail the numerical framework used to model gamma-ray emission from the solar disk. First, we construct the total magnetic field by superimposing the Alfv\'{e}n wave turbulence result onto the background magnetostatic flux structure. Our subsequent simulation is then factorized into a two-stage process, where the calculation of GCR trajectories is decoupled from the physics of individual $pp$ interactions.

This is possible because the probability of an ingoing GCR proton producing an outgoing gamma ray is small. See Figure~7 in Paper~\citetalias{2024ApJ...961..167L}, where we show that the interaction probability of relevant proton GCRs in our model is no more than $13\%$. This means that the probability of a relevant GCR proton interacting twice or more is very small. The underlying reason is the exponential variation of the atmospheric density with depth.

In the first stage, we simulate the trajectories of GCR protons as they propagate through the network field, flux tubes, and intergranular sheets. In the second stage, we compute the total gamma-ray emission by integrating the gamma-ray flux from $pp$ interactions along these pre-computed trajectories. Finally, the contributions from helium in both the GCR and the solar gas are incorporated using a nuclear enhancement factor.

\subsection{Proton GCR Equation of Motion}

The motion of proton GCRs is described by the relativistic Lorentz force equation:
\begin{equation}
    \frac{d \boldsymbol{v}}{dt} = \frac{e}{\Gamma\left( \lvert \boldsymbol{v} \rvert \right) m_p \, c} \, \boldsymbol{v} \times \boldsymbol{B} \left(\boldsymbol{r}\right),
    \label{eq: Lorentz force equation}
\end{equation}
where $\boldsymbol{v}$ is proton velocity, $e$ is the elementary charge, $\Gamma$ is the Lorentz factor, $m_p$ is the proton rest mass, $c$ is the speed of light, and $\boldsymbol{B} \left(\boldsymbol{r}\right)$ is the total magnetic field at location $\boldsymbol{r} = (x, y, z)$ in the magnetic flux structure. Here, $\boldsymbol{r}$ is treated as a continuous spatial coordinate rather than being restricted to mesh points (see Section~\ref{subsec: numerical magnetostatic}). The electric field component is considered negligible in this context. This is because the convective flow speed ($\sim 1$~km/s) at the photosphere is five orders of magnitude smaller than the proton GCR speed ($\lvert \boldsymbol{v} \rvert \approx c$), as discussed in Paper~\citetalias{2024ApJ...961..167L}. To solve this equation of motion numerically, we employ the relativistic Boris-C solver presented in \citet{2018PhPl...25k2110Z}. This solver is equivalent to the Boris solver in \citet{Boris1970} with a gyrophase correction and conservation of the particle phase-space density.

The total magnetic field in Equation~\eqref{eq: Lorentz force equation} is taken to have the form $\boldsymbol{B} \left(\boldsymbol{r}\right) = \boldsymbol{B}_{0} \left(\boldsymbol{r}\right) + \delta \boldsymbol{B} \left(\boldsymbol{r}\right)$, where $\boldsymbol{B}_{0} \left(\boldsymbol{r}\right)$ is the magnetic field from the magnetostatic calculation in Section~\ref{sec: magnetostatic model}, and $\delta \boldsymbol{B} \left(\boldsymbol{r}\right)$ is the 3D fluctuating field. To transform the magnetic power spectrum $P_B$ from Alfv\'{e}n wave turbulence in Section~\ref{sec: Alfven wave model} to the 3D fluctuating fields, we adopt the method in \citet{1982tht..book.....B} and \citet{1999ApJ...520..204G}, in which the fluctuating field is constructed from a superposition of $N_m$ plane-wave modes. In its general form, $\delta \boldsymbol{B}$ is expressed as
\begin{equation}
    \delta \boldsymbol{B} \left(\boldsymbol{r}\right) = \sum_{n=1}^{N_m} \sqrt{P_B \left(k_n, z\right)} \, \boldsymbol{\hat{\xi}}_n \exp \left(i k_n z_n^\prime + i \beta_n\right),
    \label{eq: plane_wave_modes}
\end{equation}
where
\begin{equation}
\begin{aligned}
    \boldsymbol{\hat{\xi}}_n &= \boldsymbol{\hat{x}} \left( \cos \alpha_n \cos \theta_n \cos \phi_n - i \sin \alpha_n \sin\phi_n \right) \\
    &+ \boldsymbol{\hat{y}} \left( \cos \alpha_n \cos \theta_n \sin \phi_n + i \sin \alpha_n \cos \phi_n \right) \\
    &+ \boldsymbol{\hat{z}} \left( -\cos \alpha_n \sin \theta_n \right),
\end{aligned}
\end{equation} 
\begin{equation}
    z_n^\prime = {\sin \theta_n \cos \phi_n} x + {\sin \theta_n \sin \phi_n} y + {\cos \theta_n} z .
\end{equation}
Here, the subscript $n$ denotes the $k_n$ wave mode, with $\sqrt{P_B \left(k_n, z\right)}$ the amplitude at the height $z$, $\boldsymbol{\hat{\xi}}_n$ the polarization vector, $\alpha_n$ the polarization angle, and $\beta_n$ the phase. The propagation direction of the $k_n$ mode is described by the spherical polar coordinates $\left(\theta_n, \phi_n\right)$.

Because the Alfv\'{e}n wave turbulence calculation is performed on a thin field line compared to the magnetostatic structure in Section~\ref{sec: magnetostatic model}, we must make certain assumptions on the parameters when we map the Alfv\'{e}n wave turbulence result to the arbitrary location inside the magnetic network field, flux tube, and flux sheet structures. First, for any location at height $z$ inside the flux structure, we assume the magnetic power spectrum $P_B$ to be equal to the value $P_B \left(k_n, z\right)$ obtained from our RMHD calculation. Because the RMHD calculation is based on the footpoint rms motion at $z=0$~km and the wave propagation upward, we do not know the turbulence level below $z=0$~km. Here, we make an assumption that the turbulence level in $-200~\mathrm{km} \leq z < 0~\mathrm{km}$ is the same as that at $z=0$~km. Second, the spherical polar coordinates $\left(\theta_n, \phi_n\right)$ for each $k_n$ mode at location $\boldsymbol{r}$ are taken to be the spherical polar coordinates $\left(\theta, \phi\right)$ of $\boldsymbol{B}_{0} \left(\boldsymbol{r}\right)$. Third, we take $\alpha_n = \pi / 2$ so that $\boldsymbol{\hat{\xi}}_n$ does not have the azimuthal component in the tube-like magnetostatic structure and does not have the $x$-component in the sheet-like structure. The only random parameter is the phase $\beta_n$, which is independently assigned to each mode $k_n$ from a uniform distribution over $\beta_n$ values with $0 \leq \beta_n \leq 2 \pi$. We note that $\boldsymbol{B}_{0} \cdot \delta \boldsymbol{B} = 0$, consistent with the assumption of the RMHD approximation \citep{2011ApJ...736....3V}.

For the Network F2 case, we generated a set of 20 turbulence ensembles to model the stochastic turbulent condition; for each of the Network F1, F3-F7 cases, we generate a set of 8 turbulence ensembles. Each ensemble is defined by $N_m = 9$ random phases of $\beta_n$, with each of the 9 phases corresponding to one of the 9 bins of the power spectrum $P_B$; the highest wavenumber bin spans the dimensionless range $a_\perp$ from~18 to~20 (see Figure~\ref{fig:f2_PSD}). For each of these realizations, we simulated the particle propagation and subsequent gamma-ray emission. The resulting gamma-ray spectrum presented throughout this paper denotes the statistical average over turbulence ensembles.

We set the highest $a_\perp$ bin to 18--20 and do not extend to higher wavenumbers. Across the height range considered, the magnetic fluctuation amplitude $\sqrt{P_B} / B_0$ at $a_\perp = 20$ is only $\sim 10^{-3} - 10^{-2}$ (see Figure~\ref{fig:f2_PSD}). For larger $a_\perp$ from~20 to~40, this amplitude is $\sim 10^{-4} - 10^{-3}$, implying perturbations too weak to produce particle trajectory deflections comparable to those from magnetic mirroring within our computational domain. Extending the spectrum to higher $a_\perp$ would therefore not affect GCR propagation. Consequently, our model does not impose a cutoff in GCR kinetic energy or in the resulting gamma-ray spectrum at the scale where the GCR gyro-radius equals the smallest Alfv\'{e}n wave turbulence scale.

\subsection{GCR Injection into the Network Field, Flux Tube, and Intergranular Sheet}

In this study, we evaluate seven Network Field Line cases (F1--F7 in Figure~\ref{fig:fieldline_visualization}). In each of the seven Network Field Line cases, we assume that the properties of that single field line apply uniformly across the entire photosphere. This approach enables us to model the propagation of GCRs and the subsequent gamma-ray emission within a single magnetic structure composed of a network field, a flux tube, and an intergranular sheet.

For each of the Network Field Line case, our proton GCR injection procedure consists of the following three stages:
\begin{enumerate}
    \item All proton GCRs are first injected into the network field structure at $z=12$~Mm with an isotropic distribution. We calculate the angular and energy distributions of proton GCRs reaching the network element and those escaping the network field.
    \item We calculate the angular and energy distributions of proton GCRs reaching the network element. In a parallel simulation, we inject proton GCRs isotropically at all energies into the flux tube at $z = 700$~km and track their trajectories. These two results are then combined in the gamma-ray emission calculation to weight the angular- and energy-dependent emission from the flux tube.
    \item We calculate the angular and energy distributions of those GCRs escaping the network field. In a parallel simulation, proton GCRs are injected isotropically into the intergranular sheet at $z = 600$~km. These two results are then combined in the gamma-ray emission calculation to weight the angular- and energy-dependent emission from the intergranular sheet.
\end{enumerate}
Below, we detail the numerical implementation of this injection method.

\subsubsection{Injection into the Network Field}

Proton GCRs are injected into the network field uniformly along the $y$-axis at $z=12$~Mm. Because of the azimuthal symmetry, we only need to consider one of the two horizontal axes for injection. The injection sites begin at $y_0 = 0$~km and extend to the network field's radial boundary in increments of $\Delta y_0 = 300$~km. 

The isotropic injection procedure at each injection location $y_0$ is identical to the methodology in Paper~\citetalias{2024ApJ...961..167L}. Here we briefly describe it. The initial spherical polar coordinates of proton GCR upon injection are defined as $\left( \theta_0, \phi_0 \right)$. We consider only the particles propagating downward with the $\theta_0$ ranging from $90^\circ$ to $180^\circ$ with an increment of $\Delta \theta_0 = 1^\circ$. The azimuthal angle $\phi_0$ covers the complete range from $0^\circ$ to $360^\circ$ with an increment of $\Delta \phi_0 = 90^\circ$. For every combination of $\left(y_0, \theta_0, \phi_0\right)$, we consider $E_p^\mathrm{k}$ ranging from $3.65 \times 10^{-1}$~GeV to $3.65 \times 10^5$~GeV, with eight equally spaced logarithmic $E_p^\mathrm{k}$ bins per decade.

For each set of initial parameters $\left(y_0, \theta_0, \phi_0, E_p^\mathrm{k}\right)$, a single proton is injected into the network field, and its 3D trajectory is calculated by numerically solving the Lorentz force equation~\eqref{eq: Lorentz force equation}. A specific boundary condition is applied for $z \geq 4$~Mm, where adjacent network fields are modeled as a continuous, merged structure (see the schematic diagram in Figure~\ref{fig:schematic}). To implement this repeating boundary condition in our simulation, particles that cross the network boundary above this height are reinjected at the corresponding point on the opposite side of the domain (i.e., $x \rightarrow -x$ and $y \rightarrow -y$). The particle's velocity vector remains unchanged during this reinjection process, such that the angle of incidence is equal to the angle of exit relative to the normal direction of the boundary surface. 

Below $z = 4$~Mm, there are two criteria for how each of the particle simulations is terminated. The first criterion is when particles reach the base of the network field at $z=700$~km. These particles are then injected into the flux tubes of the network element. The second criterion is when particles pass through the network boundary surface between $z=700$~km and $z=4$~Mm. These particles are then injected into the intergranular sheets in the internetwork region (INR). We refer to both cases as ``exiting the network field.''

For both criteria, we record the number of GCRs, $Q$, and their polar exit angles, $\theta^\prime$ (defined relative to $\boldsymbol{\hat{z}}$) as they exit the network field. (For example, $\theta^\prime = 90^\circ$ is horizontal, and $\theta^\prime = 180^\circ$ is vertically downward.) The azimuthal exit angles are not considered further due to the assumption of subsequent isotropic injection into the flux tube and into the intergranular sheets. The angular and energy efficiency, $\langle f \left( \theta^\prime, E_p^\mathrm{k} \right) \rangle$, for GCRs injected either into the flux tube or the intergranular sheet is calculated using the same formulation as in Paper~\citetalias{2024ApJ...961..167L}:
\begin{equation}
    \langle f \left( \theta^\prime, E_p^\mathrm{k} \right) \rangle \equiv \frac{\Delta \phi_0}{360^\circ} \int_0^{R_\mathrm{top}} Q\left(y_0, \theta^\prime, E_p^\mathrm{k}\right) \, \frac{2 \pi y_0}{A_\mathrm{tot}} dy_0,
    \label{eq: angular_energy_efficiency}
\end{equation}
where $R_\mathrm{top}$ is the network radius at the injection height $z=12$~Mm, and $A_\mathrm{tot} = \pi R_\mathrm{top}^2$ is the total horizontal cross-section area of the network structure. The particle count $Q$ is weighted by the fractional area of each injection ring ($2 \pi  y_0  dy_0 / A_\mathrm{tot}$) to account for the larger number of protons originating from greater radii.

\begin{figure*}[t]
    \centering
    \includegraphics[width=2.12 \columnwidth]{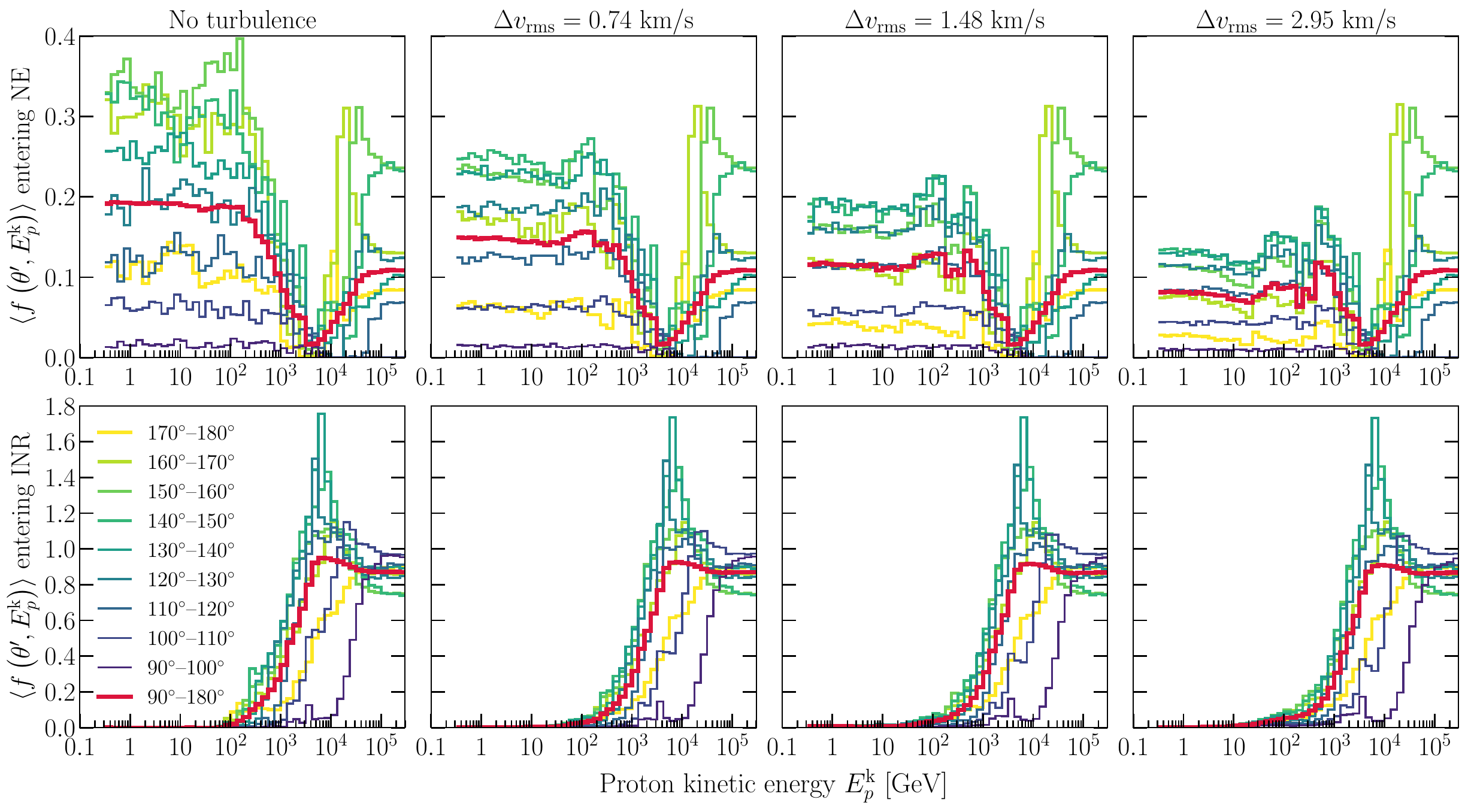}
    \caption{Angular and energy efficiency, $\langle f\left(\theta_\mathrm{exit}, E_p^\mathrm{k}\right) \rangle$, of GCRs within Network F2, shown for a no-turbulence condition and three distinct $\Delta v_\mathrm{rms}$ conditions. The top row displays this fraction for GCRs entering the network element (``NE''), while the bottom row shows it for GCRs entering the internetwork region (``INR''). The color bar indicates the GCR's exit angle from the network field.}
    \label{fig:f2_theta_exit}
\end{figure*}

Figure~\ref{fig:f2_theta_exit} presents the calculated $\langle f \left( \theta^\prime, E_p^\mathrm{k} \right) \rangle$ for the Network F2 case under four different Alfv\'{e}n wave turbulence conditions: no turbulence ($\Delta v_\mathrm{rms} = 0$~km/s), and three cases with increasing turbulence levels $\Delta v_\mathrm{rms} = 0.74$, $1.48$, and $2.95$~km/s. In each panel, the thin color lines represent $\langle f \left( \theta^\prime, E_p^\mathrm{k} \right) \rangle$ for specific GCR exit angles $\theta^\prime$ upon exiting the network field. The red thick line is the average of $\langle f \left( \theta^\prime, E_p^\mathrm{k} \right) \rangle$ over $\theta^\prime$ ranging from $90^\circ$ to $180^\circ$. Comparing the four plots in the top row, we see that a stronger turbulence condition decreases $\langle f \left( \theta^\prime, E_p^\mathrm{k} \right) \rangle$. This indicates that fewer particles successfully reach $z=700$~km as turbulence increases. Conversely, the bottom row, which shows GCRs entering the intergranular sheets, reveals very little dependence on the turbulence level. This is due to the fact that only higher-energy particles ($E_p^\mathrm{k} \gtrsim \mathrm{few} \times 10^2$~GeV, based on the plots at the bottom row) can penetrate the network field between $z=700$~km and $z=4$~Mm and enter the internetwork region. The propagation of these high-energy GCRs is barely affected by the Alfv\'{e}n wave turbulence in the height range considered in this study.

\subsubsection{Injection into the Flux Tube}

Proton GCRs are injected isotropically into the flux tube at $z=700$~km along the $y$-axis, with the injection sites, $y_0$, located at distances of $1/6$, $ 3/6$, and $5/6$ of the flux tube radius, $R_\mathrm{top}^\mathrm{tube}$, at $z=700$~km. Note that this radius varies depending on the specific network field under investigation. The initial kinetic energy and polar coordinates of the injected particles are identical to those used in the network field simulation. The trajectory simulation of each particle is terminated when the particle moves outside the vertical domain of $-200~\mathrm{km} \leq z \leq 700~\mathrm{km}$ or the horizontal domain of $-1500~\mathrm{km} \leq x, y \leq 1500~\mathrm{km}$.

\subsubsection{Injection into the Intergranular Sheet}

Proton GCRs are injected isotropically into the intergranular sheet at $z=600$~km along the $y$-axis. The injection locations are $y_0 = 50$~km, $150$~km, and $250$~km. Because $\boldsymbol{B}$ is invariant in the $x$ direction, injection is only required in the $y$ direction. The initial kinetic energy and polar coordinates of the injected particles are identical to those used in the network field and flux tube simulations. The trajectory simulation of each particle is terminated when the particle moves outside the vertical domain of $-200~\mathrm{km} \leq z \leq 600~\mathrm{km}$ or the horizontal domain of $-1500~\mathrm{km} \leq x, y \leq 1500~\mathrm{km}$.

Outside the flux boundaries of the flux tube and intergranular sheet, the magnetic field is treated as zero. Particles crossing the flux boundary follow straight-line trajectories, which we continue to track until they leave the vertical or horizontal domain. This ensures that any potential gamma-ray production in the field-free region is included in our calculations.

\subsection{Gamma-Ray Emission From \textit{pp} Collisions} \label{sec:gamma-emission pp collision}

In this subsection, we outline the methodology used to calculate the solar-disk gamma-ray spectra originating from both the flux tube and the intergranular sheet. Our primary approach is based on Section~5.3 of Paper~\citetalias{2024ApJ...961..167L}, with a key modification to the input data for gamma-ray production. In this paper, we utilize the more recent data on gamma-ray energy spectra from $pp$ collision provided in \citet{2014PhRvD..90l3014K}. This updated dataset enables us to extend the calculated disk emission spectrum to the $E_\gamma = 0.1$~GeV to $3$~GeV range. This was not previously possible in Paper~\citetalias{2024ApJ...961..167L}, which was based on the dataset of \citet{2006PhRvD..74c4018K}.

Following the framework in \citet{2014PhRvD..90l3014K}, the number of gamma-ray photons per $pp$ interaction in the $E_\gamma$ interval of $[E_\gamma, E_\gamma + dE_\gamma]$ can be expressed as
\begin{equation}
    dn_\gamma \equiv \frac{1}{\sigma_{pp}\left(E_p^\mathrm{k}\right)} \times \frac{d\sigma_\gamma}{dE_\gamma}\left(E_p^\mathrm{k}, E_\gamma \right) \times dE_\gamma.
\end{equation}
Here, $\sigma_{pp}$ is the $pp$ total inelastic cross-section and ${d\sigma_\gamma} / {dE_\gamma}$ is the gamma-ray differential cross-section from $pp$ interaction. In this work, we use the parametrized results of $\sigma_{pp}$ and ${d\sigma_\gamma} / {dE_\gamma}$ provided in Equations~(1) and~(8) of \citet{2014PhRvD..90l3014K}.

Considering a proton GCR injected into the flux tube (``tb'') or intergranular sheet (``sh'') with the initial parameters $\left( y_0, \theta^\prime, \phi^\prime, E_p^\mathrm{k} \right)$, the gamma-ray flux evaluated at the photospheric surface, integrated over the whole Sun, in the $E_\gamma$ interval of $[E_\gamma, E_\gamma + dE_\gamma]$ is expressed as
\begin{equation}
\begin{aligned}
    \left. \frac{dN_\mathrm{\gamma,\, site}}{dE_\gamma} \right\rvert_{R_\odot} &= \int_{\Omega_0} \int_{E_\gamma}^{100 E_\gamma} \frac{ \Phi_p\left(E_p^\mathrm{k}\right)}{\sigma_{pp} \left(E_p^\mathrm{k}\right)} \frac{d\sigma_\gamma}{dE_\gamma} \\
    &\times  \langle f\left(\theta^\prime, E_p^\mathrm{k}\right) \rangle_\mathrm{site} \, \cos \theta^\prime \\
    &\times \mathcal{S}_p \left( y_0, \theta^\prime, \phi^\prime, E_p^\mathrm{k} \right) \varepsilon_\mathrm{M} \left(E_p^\mathrm{k}\right) \, d E_p^\mathrm{k} d\Omega^\prime ,
    \label{eq: emission_equation}
\end{aligned}
\end{equation}
where the subscript ``site'' refers to the emission from ``tb'' or ``sh.'' Here, ${dN_\mathrm{\gamma,\, site}}/{dE_\gamma}|_{R_\odot}$ is a function of $y_0$ and $E_\gamma$. The function $\Phi_p\left(E_p^\mathrm{k}\right)$ is the proton GCR differential flux per steradian per energy interval of $[E_p^\mathrm{k}, E_p^\mathrm{k} + dE_p^\mathrm{k}]$. The component $\langle f\left(\theta^\prime, E_p^\mathrm{k}\right) \rangle_\mathrm{site}$ is the angular and energy efficiency function. Each network Field Line case has its own result of $\langle f\left(\theta^\prime, E_p^\mathrm{k}\right) \rangle_\mathrm{site}$; see Figure~\ref{fig:f2_theta_exit} for the Network F2 case. The factor $\cos\theta^\prime$ takes into account the effective number of particles entering the horizontal cross-section surface at the injection height of the flux tube or intergranular sheet. 

The function $\mathcal{S}_p$, which is identical to that in Paper~\citetalias{2024ApJ...961..167L}, represents the absorption probability of proton GCR integrated along its helical trajectory. To isolate the gamma rays that can transmit through the solar gas, we weight $\mathcal{S}_p$ by the gamma-ray transmission probability, $\zeta \left(\boldsymbol{r}\right)$, for a photon produced at $\boldsymbol{r}$. We have assumed that the gamma-ray emission is collinear with the parent GCR particle. (See \citet{2025PhRvD.112d3009G} for the discussion of noncollinear emission for $E_\gamma \lesssim \mathrm{few}$~GeV. See also Appendix~C of Paper~\citetalias{2024ApJ...961..167L}, which shows that this effect can increase the overall flux by $\mathcal{O}\left(10\%\right)$ assuming all patches of the solar surface are identical.) The function $\mathcal{S}_p$ is expressed as
\begin{equation}
    \mathcal{S}_p \left( y_0, \theta^\prime, \phi^\prime, E_p^\mathrm{k} \right) = \int_0^{\chi_p^\mathrm{max}} \zeta \left(\boldsymbol{r}\right) \frac{dP_\mathrm{abs}\left(\chi_p, E_p^\mathrm{k}\right)}{d\chi_p} d\chi_p.
    \label{eq: S_p}
\end{equation}
Here, $\chi_p = \chi_p \left(\boldsymbol{s}_p\right)$ is the integrated column density of proton GCR along the particle's 3D trajectory, $\boldsymbol{s}_p$, from the injection site at $\boldsymbol{r}_\mathrm{inj}$ to a location $\boldsymbol{r}$ where a gamma ray is produced,
\begin{equation}
    \chi_p \left(\boldsymbol{s}_p\right) = \int_{\boldsymbol{s}_p \left( \boldsymbol{r}_\mathrm{inj} \right) }^{\boldsymbol{s}_p \left( \boldsymbol{r} \right)} n_\mathrm{H} \left(z^\prime\right) \lvert d\boldsymbol{s}_p^\prime \rvert.
\end{equation}
In Equation~\eqref{eq: S_p}, $\chi_p^\mathrm{max}$ is the total column density of the particle from $\boldsymbol{r}_\mathrm{inj}$ to the exit point ($\boldsymbol{r}_\mathrm{exit}$), i.e., $\chi_p^\mathrm{max} = \chi_p \left(\boldsymbol{r}_\mathrm{exit}\right)$. The function $dP_\mathrm{abs}\left(\chi_p, E_p^\mathrm{k}\right)$ is the absorption probability for a proton GCR with kinetic energy $E_p^\mathrm{k}$ that accumulates a column density of $\chi_p$,
\begin{equation}
    P_\mathrm{abs}\left(\chi_p, E_p^\mathrm{k}\right) = 1 - \exp\bigg[ -\chi_p \sigma_{pp}\left(E_p^\mathrm{k}\right)\bigg].
    \label{eq: absorption probability}
\end{equation}
The gamma-ray transmission probability $\zeta\left(\boldsymbol{r}\right)$ describing the probability for a gamma-ray photon produced at $\boldsymbol{r}$ to be transmitted through the solar gas is expressed as
\begin{equation}
    \zeta\left(\boldsymbol{r}\right) = \exp \left( - \frac{t_\gamma}{\lambda_\gamma}\right).
\end{equation}
Here, the function $t_\gamma \left(\boldsymbol{r}\right)$ is the integrated mass column density along the emitted gamma-ray photon path, from the emission site ($\boldsymbol{r}$) to a far distance from the Sun ($\boldsymbol{r}_\mathrm{infty}$),
\begin{equation}
    t_\gamma \left( \boldsymbol{r} \right) = \int_{\boldsymbol{r}}^{\boldsymbol{r}_\mathrm{infty}} \rho\left(\boldsymbol{r}^\prime\right) \lvert d\boldsymbol{r}^\prime \rvert.
\end{equation}
The quantity $\lambda_\gamma$ represents the photon mass attenuation length. For gamma rays with $E_\gamma \gtrsim 0.1$~GeV in hydrogen and helium gas, the attenuation length is approximately constant at $\lambda_\gamma \approx 80~\mathrm{g/cm^2}$ \citep[][pg.~558]{2022PTEP.2022h3C01W}.

The final term, $\varepsilon_\mathrm{M} \left(E_p^\mathrm{k}\right)$, in Equation~\eqref{eq: emission_equation} is the nuclear enhancement factor. It accounts for the additional gamma-ray yield from interactions involving helium nuclei present in the cosmic rays and the solar gas, supplementing the primary yield from $pp$ collision. For $E_p^\mathrm{k} \geq 1$~GeV, we adopt the ``{Tripathi(1999)+modified}'' dataset from Figure~14 of \citet{2014PhRvD..90l3014K}, which builds upon the results of \citet{1999NIMPB.155..349T} below 1~TeV and is modified to include an energy-dependent cross-section at higher energies. However, neither study provides this factor for the range below 1~GeV. Therefore, in this low-energy regime below 1~GeV, we assume a constant value equal to the factor at 1~GeV, i.e., $\varepsilon_\mathrm{M} = \varepsilon_\mathrm{M} \left( 1~\mathrm{GeV} \right) = 1.99$.

The gamma-ray flux in Equation~\eqref{eq: emission_equation} applies to one injection site at $y=y_0$. To account for the horizontal cross-section area of the flux tube and intergranular sheet at their injection heights, we weight Equation~\eqref{eq: emission_equation} over the area. For the injection into the flux tube case, the averaged gamma-ray flux is expressed as
\begin{equation}
    \left. \frac{d\overline{N}_\mathrm{\gamma,\, tb}}{dE_\gamma} \right\rvert_{R_\odot} = \int_0^{R_\mathrm{top}} \left. \frac{dN_\mathrm{\gamma,\, tb}}{dE_\gamma} \right\rvert_{R_\odot} \frac{2 \pi y_0}{A_\mathrm{tot}} dy_0,
    \label{eq: avg tb emission}
\end{equation}
where $A_\mathrm{tot} = \pi R_\mathrm{top}^2$, with $R_\mathrm{top}$ being the flux tube radius at $z=700$~km. For the injection into the intergranular sheet, it is expressed as
\begin{equation}
    \left. \frac{d\overline{N}_\mathrm{sh}}{dE_\gamma} \right\rvert_{R_\odot} = \int_0^{y_\mathrm{top}} \left. \frac{dN_\mathrm{\gamma,\, sh}}{dE_\gamma} \right\rvert_{R_\odot} \frac{y_0}{y_\mathrm{top}} dy_0,
    \label{eq: avg sh emission}
\end{equation}
where $y_\mathrm{top}$ is the half-width of the intergranular sheet at $z=600$~km.

The total gamma-ray flux evaluated at the solar surface is the sum of the contribution from the flux tube in Equation~\eqref{eq: avg tb emission} and the contribution from the intergranular sheet in Equation~\eqref{eq: avg tb emission}. At 1~au from the Sun, the total gamma-ray flux, ${dN_\gamma}/{dE_\gamma}$, becomes
\begin{equation}
    \frac{dN_\gamma}{dE_\gamma} = \left(\frac{R_\odot}{1~\mathrm{au}}\right)^2 \left( \left. \frac{d\overline{N}_\mathrm{tb}}{dE_\gamma} \right\rvert_{R_\odot} + \left. \frac{d\overline{N}_\mathrm{sh}}{dE_\gamma} \right\rvert_{R_\odot} \right).
\end{equation}


\section{Numerical Results For The Predicted Gamma-Ray Spectrum} \label{sec: results}

In this section, we present our numerical results for the predicted gamma-ray spectra, emission angles, and emission heights. First, in Sections~\ref{subsec: spectra line 2} and \ref{subsec: angle height line 2}, we analyze the Network F2 case, comparing the scenario with no turbulence to three cases with increasing turbulence levels ($\Delta v_\mathrm{rms} = 0.74,\, 1.48,\, 2.95$~km/s). Then, in Section~\ref{subsec: all field lines}, we compare the spectra for all seven Network Field Line cases under a standard turbulence level of $\Delta v_\mathrm{rms} = 1.48$~km/s.

\subsection{Gamma-Ray Spectra for Network F2 Case} \label{subsec: spectra line 2}

\begin{figure}[t]
   \centering
   \includegraphics[width=1.0 \columnwidth]{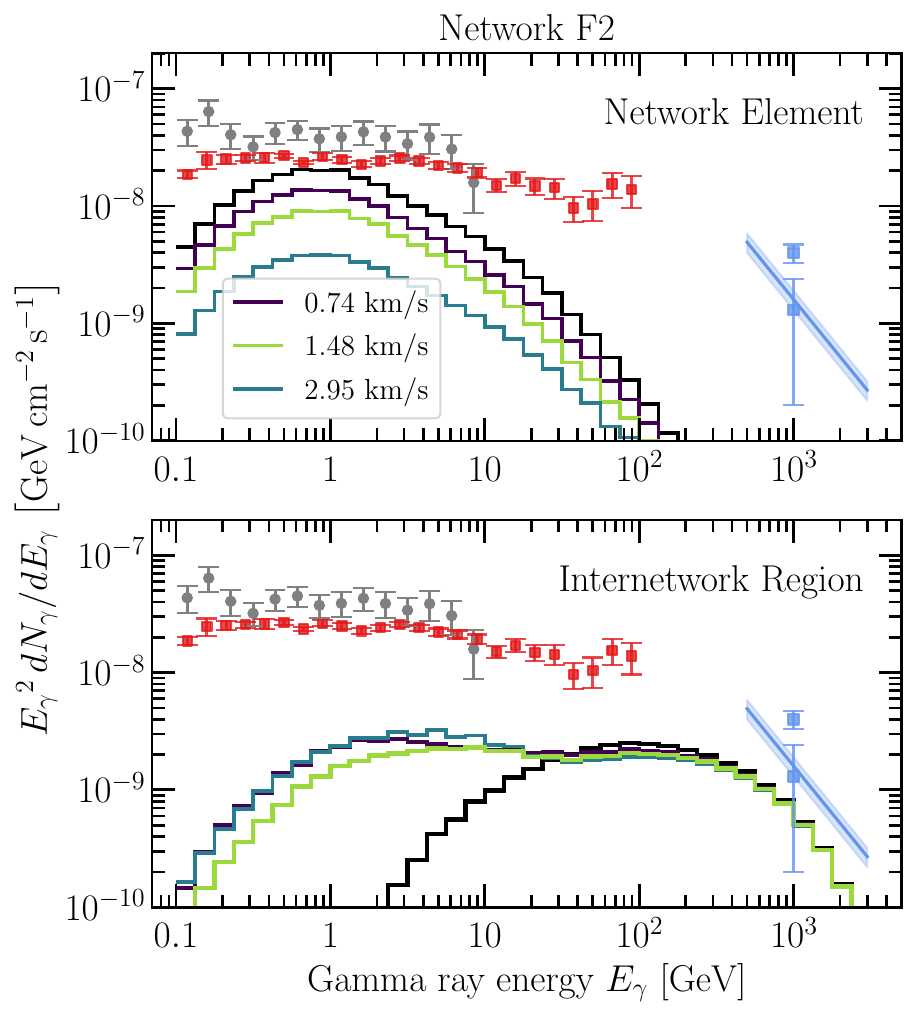}
   \caption{Gamma-ray emission contributions from the network element (top) and internetwork region (bottom), for Network F2. Lines are results for the no-turbulence condition (black) and three $\Delta v_\mathrm{rms}$ conditions (colors). Observational data are identical to those in Figure~\ref{fig:f2_gamma_combine}.}
   \label{fig:f2_gamma_NE_INR}
\end{figure}

\begin{figure*}[t]
    \centering
    \includegraphics[width=2.12 \columnwidth]{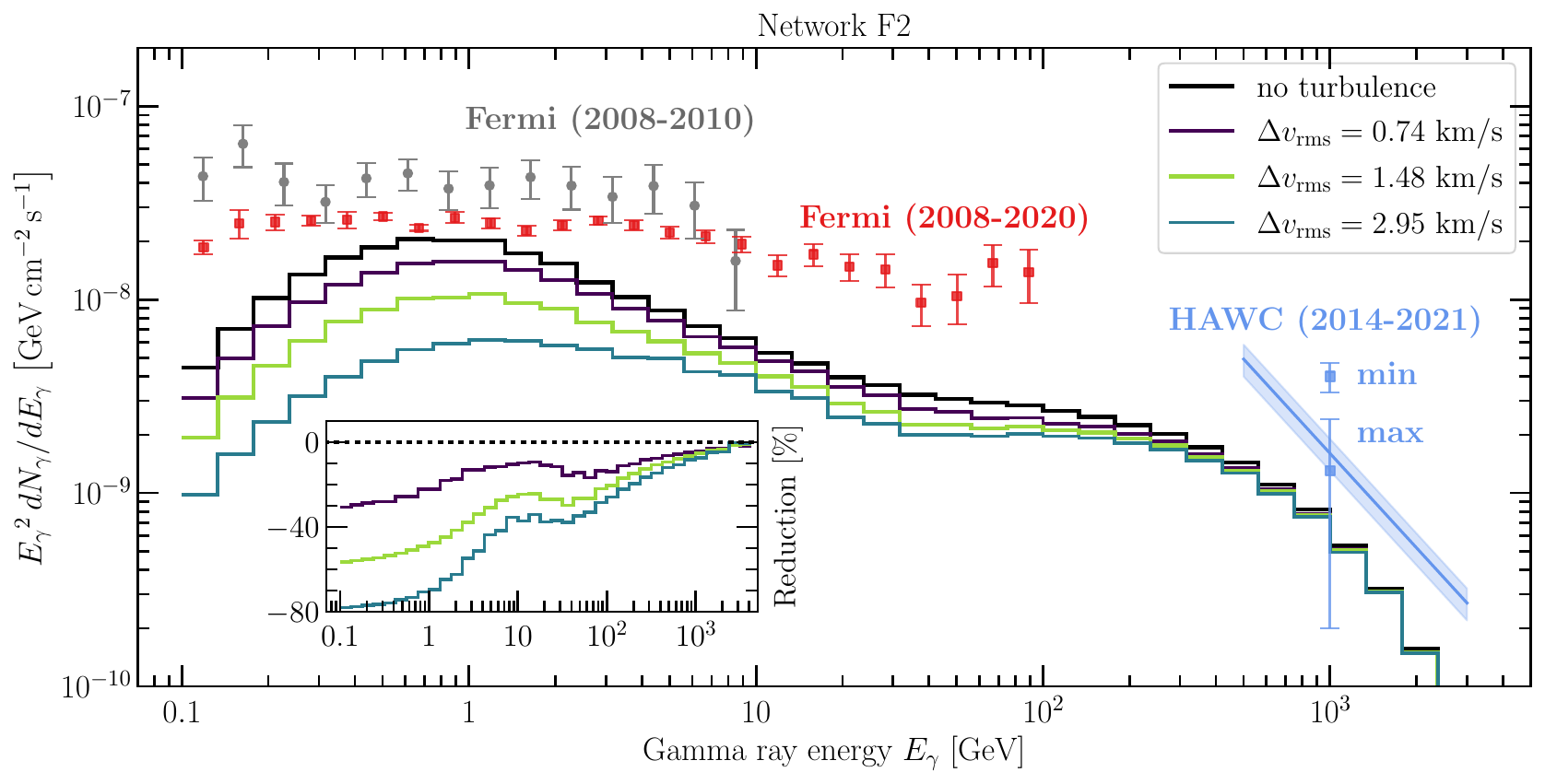}
    \caption{Gamma-ray spectrum of the solar disk for Network F2. Calculated gamma-ray spectra (lines) are presented for a no-turbulence condition (black) and three $\Delta v_\mathrm{rms}$ conditions (colors). Inset: Flux reduction for the three $\Delta v_\mathrm{rms}$ conditions relative to the no-turbulence condition. Observations presented are: Fermi-LAT data from solar minimum \citep[2008--2010;][]{2011ApJ...734..116A} and the full solar cycle \citep[2008--2020;][]{2022PhRvD.105f3013L}; and HAWC data covering solar maximum (``max,'' 2014--2017) and minimum \citep[``min,'' 2018--2021;][]{2023PhRvL.131e1201A}. The HAWC 6.1-year data (2014--2021) in the 0.5--2.6~TeV energy bin is shown by a blue best-fit line with the shaded band indicating its statistical uncertainties.}
    \label{fig:f2_gamma_combine}
\end{figure*}

Figure~\ref{fig:f2_gamma_NE_INR} shows the gamma-ray emission for the Network F2 case with contributions from the flux tubes in the network element (top panel) and the intergranular sheets in the internetwork region (bottom panel). (Note that Paper~\citetalias{2024ApJ...961..167L} was limited to flux tubes within network elements, whereas the present work incorporates the overlying network field and the associated Alfv\'{e}n wave turbulence.) A comparison reveals that the network element dominates the emission for gamma-ray energies $E_\gamma \lesssim 10$~GeV, whereas the internetwork region becomes the dominant source for $E_\gamma \gtrsim 30$~GeV. This energy-dependent transition of the dominant emission region occurs because lower-energy GCRs ($E_p^\mathrm{k} \lesssim 100$~GeV) are magnetically confined to the network fields and flow into the flux tubes. In contrast, higher-energy GCRs ($E_p^\mathrm{k} \gtrsim 300$~GeV) are sufficiently energetic to penetrate the network magnetic structure and enter the internetwork region, as supported by the angular and energy efficiency results in Figure~\ref{fig:f2_theta_exit}. This result is broadly consistent with the findings of Paper~\citetalias{2024ApJ...961..167L}.

\begin{figure*}[t]
    \centering
    \includegraphics[width=2.12 \columnwidth]{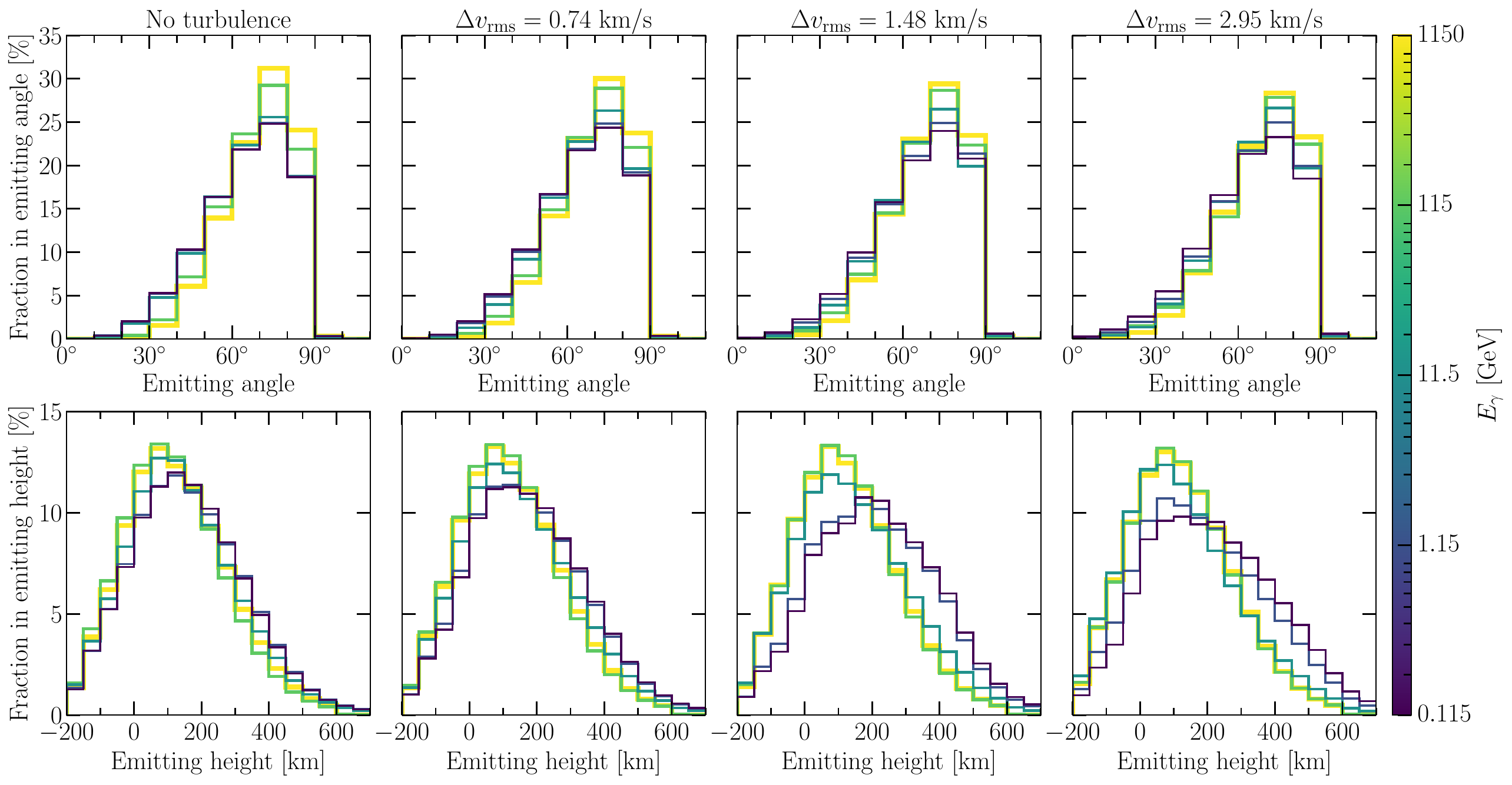}
    \caption{Emission angles and heights of gamma rays successfully transmitted from the solar disk, for Network F2. Results are shown for a no-turbulence condition and three $\Delta v_\mathrm{rms}$ conditions.}
    \label{fig:f2_emitting_angle_height}
\end{figure*}

The top panel of Figure~\ref{fig:f2_gamma_NE_INR} shows that as $\Delta v_\mathrm{rms}$ increases from~0 to~2.95~km/s, the gamma-ray flux from the network element decreases by 50--60\% across the 0.1--100~GeV energy range. This reduction occurs because turbulence reflects GCR particles at higher altitudes where the gas density is lower. To estimate this shift in altitude, we consider the absorption probability (Equation~\ref{eq: absorption probability}), which is proportional to the hydrogen number density to the leading order. As a result, a 50\% flux decrease thus implies a 50\% decrease in the hydrogen density at the reflection height. For a density scale height of approximately 100~km at the photosphere layer, this corresponds to an increase in the average GCR reflection height of $\Delta z \approx70$~km.

The bottom panel of Figure~\ref{fig:f2_gamma_NE_INR} reveals that gamma-ray emission from the internetwork region is anti-correlated with the turbulence level for energies $E_\gamma \lesssim 30$~GeV. This behavior can be attributed to the effect of turbulence on the trajectories of lower-energy GCRs near the magnetic boundary of the network field: The turbulence causes trajectory perturbations that allow some GCRs, which would otherwise be confined by the magnetic field, to cross this boundary. A higher level of turbulence therefore leads to a more efficient influx of these lower-energy GCRs into the internetwork region. For $E_\gamma \gtrsim 30$~GeV, the flux is nearly independent of the turbulence level. This latter result indicates that Alfv\'{e}n wave turbulence, as modeled here, cannot account for the observed time correlation in the 1~TeV HAWC data. A possible explanation for this discrepancy lies in the limitation of the isolated flux tube model, which neglects the complex interactions that likely occur between adjacent magnetic field lines in the solar atmosphere. By ignoring this coupling, our model likely omits the long-wavelength Alfv\'{e}n wave turbulence with which higher-energy GCRs can resonantly scatter. A more realistic framework is therefore required to accurately capture the turbulence dynamics. For instance, the model developed by \citet{2017ApJ...849...46V} explicitly includes coupling between neighboring flux tubes in the chromosphere and corona. Such interactions are critical for determining the temporal and spatial characteristics of the wave-driven turbulence and may be essential for explaining the time correlation observed by HAWC.

\begin{figure*}[t]
    \centering
    \includegraphics[width=1.65 \columnwidth]{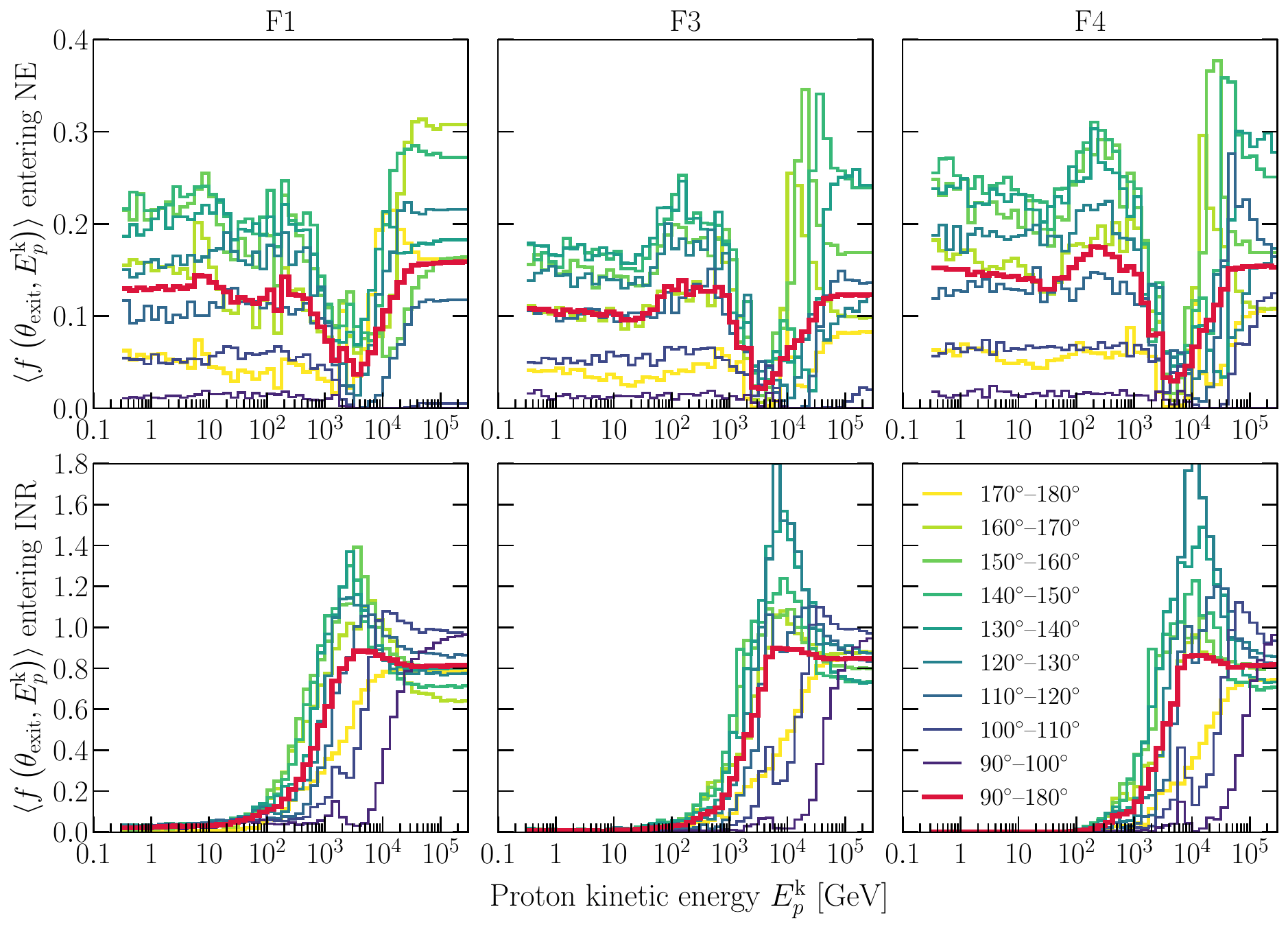}
    \caption{Angular and energy efficiency, $\langle f\left(\theta_\mathrm{exit}, E_p^\mathrm{k}\right) \rangle$, of GCRs entering the network element (top row), and the internetwork region (bottom row). Results are shown for the Network F1, F3, and F4 from the quiet photospheric region, all calculated with $\Delta v_\mathrm{rms}$ fixed at 1.48~km/s.}
    \label{fig:quiet_lines_theta_exit}
\end{figure*}

\begin{figure*}[t]
    \centering
    \includegraphics[width=1.65 \columnwidth]{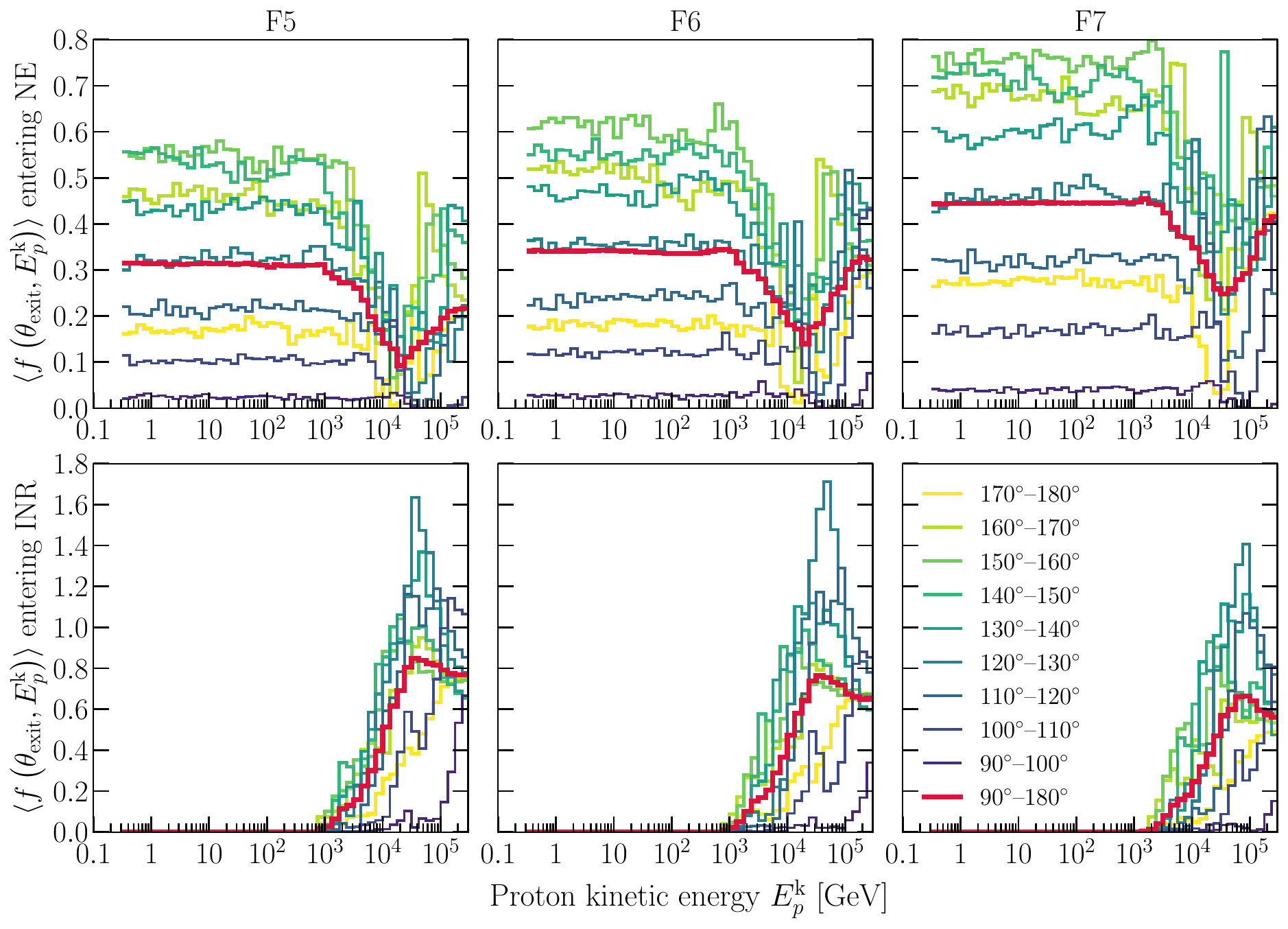}
    \caption{Same as Figure~\ref{fig:quiet_lines_theta_exit}, but for Network F5, F6, and F7 close to an active photospheric region.}
    \label{fig:active_lines_theta_exit}
\end{figure*}

Figure~\ref{fig:f2_gamma_combine} shows the total gamma-ray energy spectra for Network F2, comparing the case with no turbulence to three cases with varying $\Delta v_\mathrm{rms}$. First, the gamma-ray flux below approximately 100 GeV is suppressed, which we attribute to turbulence perturbing the particle trajectory and increasing the GCR reflection height. This suppression results in a flatter spectrum slope between 1~GeV and 100~GeV that aligns well with Fermi-LAT observations. Second, at much higher $E_\gamma$ of about 1~TeV, the spectrum steepens to $dN_\gamma /dE_\gamma \sim E_\gamma^{-3.6}$, showing good agreement with HAWC observations. This steepening occurs due to the ineffectiveness of GCR reflection for high-energy particles within the narrow (a few hundred kilometers wide) intergranular sheets. The excellent agreement of the overall spectral shape from 1~GeV to 1~TeV strongly suggests that the observed gamma-ray patterns are shaped by a variety of filamentary magnetic structures.

While our model successfully reproduces key spectral features from 1~GeV to 1~TeV, Figure~\ref{fig:f2_gamma_combine} also highlights two areas for future refinement. First, for $E_\gamma < 1$~GeV, the model predicts a spectrum slope of $dN_\gamma /dE_\gamma \sim E_\gamma^{-0.4}$, which is harder than the $dN_\gamma /dE_\gamma \sim E_\gamma^{-2}$ slope observed by Fermi-LAT. Second, the predicted flux from 0.1~GeV to 1~TeV is lower than observations by a factor of~2 to~5. We discuss the potential causes of these divergences and their implications for refining the model in Section~\ref{sec: discussion}.

\subsection{Gamma-Ray Emission Angle and Height for Network F2 Case} \label{subsec: angle height line 2}

Figure~\ref{fig:f2_emitting_angle_height} shows the angular and height distributions for gamma rays successfully transmitted from the Sun. The top panels reveal that the majority of the escaping flux is emitted at large angles, specifically between $50^\circ$ and $90^\circ$ relative to the local surface normal, with less than $10\%$ of the flux emitted within $50^\circ$. This preference for large emission angles is a consequence of magnetic mirroring being the dominant particle reflection mechanism over Alfv\'{e}n turbulence effects. In a mirroring scenario, particles are primarily reflected at their lowest heights, where their pitch angle reaches $90^\circ$ and where the ambient gas density is highest. Consequently, the resulting gamma-ray emission is directed tangentially to or slightly above the surface but not in the direction normal to the surface. The sharp cutoff in emission above $90^\circ$ is caused by atmospheric absorption: Gamma rays emitted at angles greater than approximately $93^\circ$ are fully absorbed by the solar gas before they can escape.

The bottom panel of Figure~\ref{fig:f2_emitting_angle_height} shows the distribution of emission heights for escaping gamma rays. The emission is concentrated in a layer extending from approximately $-100$~km to $400$~km, which includes the full photosphere and $100$~km into the uppermost convection zone. Our analysis reveals an energy-dependent response to turbulence in this region. As $\Delta v_\mathrm{rms}$ increases, the peak emission height for lower-energy gamma rays shifts upward from $z \approx 100$~km to $z \approx 200$~km, while the peak for higher-energy gamma rays remains insensitive to the turbulence level. This dichotomy is consistent with our findings in Figure~\ref{fig:f2_gamma_combine}, which demonstrates that the modulating effect of Alfv\'{e}n wave turbulence is confined to gamma rays with $E_\gamma \lesssim 100$~GeV.

\begin{figure*}[t]
    \centering
    \includegraphics[width=2.12 \columnwidth]{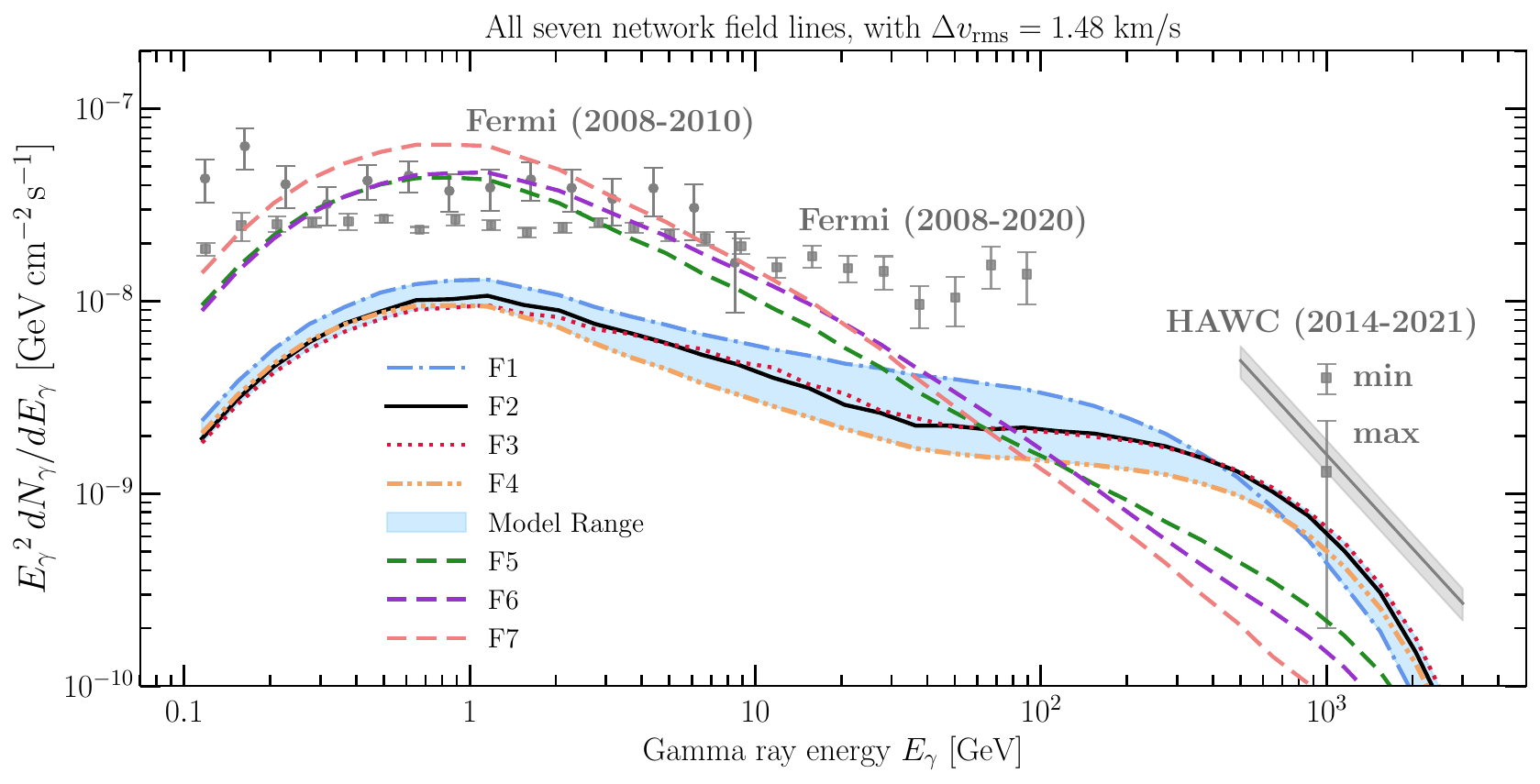}
    \caption{Gamma-ray spectra are presented for all seven network field lines, all calculated with the standard $\Delta v_\mathrm{rms}$ value of 1.48~km/s. The blue band denotes the model range from a quieter solar region networks F1--F4 shown in this work. In contrast, F5--F7 close to an active region have spectrum slopes deviate from gamma-ray observational data.}
    \label{fig:all_lines_gamma}
\end{figure*}

\subsection{All Seven Field Line Cases} \label{subsec: all field lines}

Figures~\ref{fig:quiet_lines_theta_exit} and~\ref{fig:active_lines_theta_exit} presents the calculated $\langle f\left(\theta_\mathrm{exit}, E_p^\mathrm{k}\right) \rangle$, assuming the standard turbulence of $\Delta v_\mathrm{rms} = 1.48$~km/s. Figure~\ref{fig:quiet_lines_theta_exit} corresponds to the Network F1, F3, and F4 in the quiet photospheric region, while Figure~\ref{fig:active_lines_theta_exit} presents the corresponding results for the Network F5, F6, and F7 close to the active photospheric region. (Note that the result for the F2 case is shown separately in Figure~\ref{fig:f2_theta_exit}.) As established in the HI and MDI images in Figure~\ref{fig:fieldline_visualization} and the background magnetic field strength in Figure~\ref{fig:all_lines_B0}, F1 to F4 represent quiescent solar regions with similar weaker magnetic field strength profiles. In contrast, F5 to F7 represent a more active region with a significantly stronger field. As also shown in Table~\ref{table: field_line_characteristics}, the total magnetic flux increases monotonically from configuration F1 to F7.

Figures~\ref{fig:quiet_lines_theta_exit} and~\ref{fig:active_lines_theta_exit} reveals two key trends in the angular and energy efficiency, $\langle f\left(\theta_\mathrm{exit}, E_p^\mathrm{k}\right) \rangle$, that arise from the different magnetic profiles and fluxes of the Network Field Lines. The first trend is observed in the network element (top row), where the efficiency for F5--F7 cases are substantially higher than for F1--F4 cases. This enhanced efficiency is a direct consequence of its stronger magnetic field profile and higher total magnetic flux ($\Phi_B \geq 3 \times 10^{20}$~Mx), which leads to higher confinement and more effectively confines and channels GCRs into the network element.

The second trend is observed in the internetwork region (bottom rows of Figures~\ref{fig:quiet_lines_theta_exit} and~\ref{fig:active_lines_theta_exit}). Network F1 shows a non-zero efficiency $\langle f\left(\theta_\mathrm{exit}, E_p^\mathrm{k}\right) \rangle$ for $E_p^\mathrm{k} \lesssim 100$~GeV, as its smaller magnetic structure in size is more easily penetrated. This low-energy penetration diminishes progressively for F2 through F7. As the total magnetic flux and the corresponding network size increase, the structure becomes more effective at trapping these particles. This effect is strongest in F5--F7 cases, whose high magnetic fluxes (and thus large magnetic structures in size) shield the internetwork from all GCRs of $E_p^\mathrm{k}$ below 1~TeV.

Figure~\ref{fig:all_lines_gamma} displays the predicted gamma-ray spectra for all seven Network Field Line cases. A key observation is that F1--F4 produce nearly identical gamma-ray spectra. This similarity arises from their comparable magnetic field strength profiles (Figure~\ref{fig:all_lines_B0}) and their common origin from a magnetically quiet patch of the Sun (Figure~\ref{fig:fieldline_visualization}). Accordingly, the blue band spanning the results for F1--F4 in Figure~\ref{fig:all_lines_gamma} represents our model prediction of the emission range for these quieter patches of the Sun. While the overall predicted flux of the blue band is a factor of 2--5 lower than observed data, the spectrum slope of this model shows good consistency with Fermi-LAT observations from 1~GeV to 100~GeV and with HAWC data at 1~TeV. Based on this strong agreement in the spectral shape, we propose that the observed quiescent solar gamma-ray emission likely originates from such quieter photospheric regions.

In Figure~\ref{fig:all_lines_gamma}, the spectra for F5--F7 (representing an active region) have slopes of $dN_\gamma / dE_\gamma \sim E_\gamma^{-2.7}$ from 1~GeV to 1~TeV, which is inconsistent with observations. Notably, this spectral index of $-2.7$ matches that of the incident GCRs. This correspondence is a consequence of the strong magnetic field strengths of the F5--F7 that are highly efficient at capturing GCRs of all relevant energies, thus preserving the original GCR spectrum slope inside the interaction region in the flux tube. The resulting gamma-ray spectrum thus also follows the spectral index of $-2.7$.


\section{Discussion} \label{sec: discussion}

In this section, we discuss the primary findings of this work and outline key directions for future research.

\subsection{Key Insights From This Work}

Drawing from the results in Figure~\ref{fig:all_lines_gamma}, we identify several key insights.

\begin{enumerate}

\item The hard gamma-ray spectrum ($dN_\gamma / dE_\gamma \propto E_\gamma^{-2.2}$) observed by Fermi-LAT between 1~GeV and 100~GeV is likely produced by the combined effect of multiple, finite-sized, filamentary magnetic flux structures, consistent with findings in Paper~\citetalias{2024ApJ...961..167L}.

\item The softer spectrum ($dN_\gamma / dE_\gamma \propto E_\gamma^{-3.6}$) seen by HAWC around 1~TeV must originate from those filamentary, finite-sized flux structures, which are a few hundred kilometers wide at the photospheric base. In contrast, a magnetic structure with a horizontal extent \emph{significantly} larger than a few hundred~km at the photosphere will yield only $dN_\gamma / dE_\gamma \propto E_\gamma^{-2.7}$, which is inconsistent with HAWC observations.

\item Motivated by the successful results from F1--F4, the observed gamma-ray flux likely originates from quiet photospheric regions. On the contrary, within active regions (the F5--F7 cases), GCRs either are not reflected by magnetic fields at the optimal heights ($z = -100$~km to 400~km) for efficient gamma-ray production or are largely shielded from entry.

\item Our Alfv\'{e}n wave turbulence model shows turbulence modulates the gamma-ray flux for $E_\gamma \lesssim 100$~GeV but leaves emission at higher energies unaffected. Therefore, while this model on Alfv\'{e}n wave turbulence may be a clue about the anti-correlation with the solar cycle at lower energies, our results show it is insufficient to explain the phenomenon in its entirety. An alternative or additional mechanism must be responsible for the modulation of gamma-ray flux observed for $E_\gamma \gtrsim 100$~GeV.

\end{enumerate}

\subsection{Future Directions}

Our predicted spectrum slope of Network F1--F4 aligns with Fermi-LAT and HAWC observations. This agreement suggests that our model, which incorporates filamentary field structures in quiescent photospheric regions, represents a viable gamma-ray emission mechanism. While the model is promising, several areas warrant further refinement, which we outline below as key directions for future work.

\subsubsection{Searching Other Patches of the Photosphere}
Our predicted flux underestimates the total gamma-ray flux by a factor of 2--5. This discrepancy may be explained by stronger GCR reflection than is accounted for in our model. A stronger mean field would enhance the reflection of GCRs, leading to a higher gamma-ray yield. It is plausible that other quiescent photospheric regions, which possess stronger magnetic fields than those modeled in this work (Figure~\ref{fig:all_lines_B0}), could provide this enhanced reflection and account for the observed flux. A systematic investigation of many other quiescent photospheric regions and the subsequent gamma-ray emission is an interesting direction for future work.

\subsubsection{Different $pp$ Collision Models}
Another possible explanation for our predicted underestimation of the gamma-ray flux by a factor of 2--5 may, in part, arise from discrepancies among different $pp$ collision models. \citet{2025JCAP...04..043D} showed that variations in the adopted $pp$ collision model can lead to differences in the predicted gamma-ray flux of up to a factor of~$\sim 2$. In particular, they reported that the flux predicted by \citet{2014PhRvD..90l3014K} is a factor of $\sim 3$ lower than that obtained by \citet{2006PhRvD..74c4018K}. Considering these model-dependent uncertainties, the accuracy of our solar-disk gamma-ray flux prediction is limited to within a factor of $\sim \text{2--3}$. Moreover, since our calculation uses the $pp$ collision model from \citet{2014PhRvD..90l3014K}, the predicted solar-disk gamma-ray flux presented in this work is likely an underestimate of the true flux. Consequently, the apparent discrepancy between our prediction and the observed gamma-ray data by a factor of 2--5 might be less significant than it initially appears.

\subsubsection{Modified Nuclear Enhancement Factor}
Our predicted spectrum between $E_\gamma = 0.1$~GeV and 1~GeV fails to reproduce the spectrum slope measured by Fermi-LAT in this range. From Figure~\ref{fig:all_lines_gamma}, it is clear that the emission is somehow suppressed as $E_\gamma$ decreases from 1~GeV to 0.1~GeV. This discrepancy may be due to the nuclear enhancement factor, which we adopted from \citet{2014PhRvD..90l3014K}, as this factor is only valid for $E_p^\mathrm{k} \geq 1$~GeV. At $E_p^\mathrm{k}$ below 1~GeV, interactions involving nuclei ($p$-nucleus and nucleus-nucleus) become increasingly significant relative to $pp$ collisions. For instance, while $\pi^0$ production from $pp$ interaction ceases below the kinematic threshold of $E_p^\mathrm{k} \approx 0.28$~GeV, nucleus-nucleus interactions can still produce $\pi^0$ below $E_p^\mathrm{k} = 0.28$~GeV/nucleon \citep{2014PhRvD..90l3014K}. Consequently, the true enhancement factor is likely greater than the constant value of $\varepsilon_\mathrm{M} = 1.99$ for $E_p^\mathrm{k} \leq 1~\mathrm{GeV}$ used in our gamma-ray calculation in Section~\ref{sec:gamma-emission pp collision}. To our knowledge, a detailed quantitative analysis of $\varepsilon_\mathrm{M}$ for $E_p^\mathrm{k} \leq 1~\mathrm{GeV}$ in this low-energy regime is not yet available. Further investigation using modern hadronic interaction codes, such as \texttt{FLUKA} \citep{2015_Battistoni_fluka, 2022FrP.....9..705A}, is therefore warranted to resolve this issue.

We note the recent work by \cite{2023PhRvD.107h3031O}, which provides energy-differential cross sections for gamma-ray production across a wide range of projectile energies ($0.1$ to~$10^7$~GeV/n). Their data allow for a direct calculation of the gamma-ray yield from nucleus-nucleus interactions using AMS-02 cosmic-ray composition data \citep{2016PhRvL.117w1102A, 2018PhRvL.121e1103A}, providing an alternative to the nuclear enhancement factor used here. The application of this model to solar gamma-ray emission is a promising avenue for future work.

\subsubsection{Alfv\'{e}n Wave Turbulence and the Time Variation of the Gamma-Ray Flux}
The findings in our study demonstrate the important energy-dependent role of Alfv\'{e}n wave turbulence in altering the time variation of solar gamma-ray emission. While our model demonstrates that turbulence suppresses gamma-ray fluxes for $E_\gamma \lesssim 100$~GeV by altering GCR reflections, its effect becomes negligible at higher energies. For $E_\gamma \gtrsim 100$~GeV, the gyro-radius of GCRs becomes comparable to or larger than the turbulence scales modeled in isolated flux tubes, leading to minimal pitch-angle scattering and weak modulation. However, the time-variability of the gamma-ray flux at $E_\gamma = 1$~TeV observed by HAWC cannot be explained by the isolated field line turbulence model shown in this work.

To address the modulation for $E_\gamma \gtrsim 100$~GeV, in our future studies we will move beyond isolated flux tubes and adopt a framework that accounts for coupled magnetic field lines. One promising avenue is to implement the coupled-field model developed by \citet{2017ApJ...849...46V}, which describes energy transfer and wave dynamics between adjacent field lines in the chromosphere and corona. Incorporating interacting field lines  may produce longer-wavelength Alfv\'{e}n modes that are capable of resonantly interacting hadronic GCRs with $E_p^\mathrm{k} \gtrsim 100$~GeV, thereby enabling turbulence-mediated modulation at TeV energies. Additionally, a more complete model should consider the large-scale magnetic topology and the temporal evolution of the photospheric magnetic field. These features could influence the spatial distribution and temporal variability of the gamma-ray flux and may be particularly important for interpreting the solar-cycle anti-correlation at the highest observed energies. Together, these extensions aim to provide a more comprehensive physical picture of the gamma-ray modulation mechanism in the solar atmosphere.


\section{Conclusions} \label{sec: conclusion}

In this paper, we have modeled the hadronic GCR propagation in the multi-scale, open magnetic field to understand how this structure shapes the observed solar-disk gamma-ray spectrum. We show that GCR transport is governed by two primary, energy-dependent processes: magnetic mirroring by photospheric structures and scattering from Alfv\'{e}n wave turbulence. As GCRs approach the solar surface, their transport behavior diverges with energy. From the corona down to the chromosphere, the trajectories of lower-energy GCRs are guided by the mean magnetic field of the network field but also experience pitch-angle scattering from turbulence as they enter the flux tubes of network elements. In contrast, higher-energy GCRs are less confined by the mean magnetic field of the network field and are not affected by turbulence, allowing them to penetrate more directly into intergranular sheets in the internetwork regions. Deeper in the atmosphere---at the photosphere and uppermost convection zone where gamma rays are produced via $pp$ interactions---the filamentary nature of flux tubes and intergranular sheets causes an energy-dependent magnetic mirroring efficiency on GCRs, leading to a suppression of gamma-ray emission at higher energies and thereby shaping the final observed spectrum.

Our model successfully reproduces the solar-disk gamma-ray spectrum slopes from 1~GeV to 1~TeV, demonstrating excellent agreement with Fermi-LAT and HAWC data. In the 1--100~GeV range, emission from the quiet photosphere is contributed from the filamentary field structures of both flux tubes and intergranular sheets, with the overall flux suppressed by $\lesssim 60\%$ due to Alfv\'{e}n wave turbulence. The resulting spectrum slope follows $dN_\gamma / dE_\gamma \sim E_\gamma^{-2.4}$. At 1~TeV, the spectrum softens significantly to $dN_\gamma / dE_\gamma \sim E_\gamma^{-3.6}$. This softening occurs because of the inefficient magnetic mirroring of higher-energy GCRs ($E_p^\mathrm{k} \gtrsim \mathrm{few~TeV}$) from intergranular sheets with a finite horizontal extent of only a~few hundred~km. The strong agreement between our quiet photosphere model and the observations confirms this as the origin of the emission. In contrast, models of more active regions predict a spectrum slope of $dN_\gamma / dE_\gamma \sim E_\gamma^{-2.7}$, which is inconsistent with the data.

Despite its successes, our model has several limitations that provide clear directions for future refinement. 
\begin{enumerate}
\item Our predicted gamma-ray flux from 1~GeV to 1~TeV is a factor of 2--5 lower than observations. We propose that exploring patches of the quiet photosphere with slightly higher magnetic field strengths could enhance the reflection rate and enhance the gamma-ray flux.

\item Uncertainties in the $pp$ collision models can cause variations in the predicted gamma-ray flux by a factor of~$\sim 2$, limiting the accuracy of solar-disk gamma-ray predictions to the same level. Improved hadronic cross-section modeling from the broader particle physics community is essential for enhancing the predictive power of solar-disk gamma-ray studies of solar magnetic fields.

\item The predicted spectrum slope and overall flux below 1~GeV do not match the Fermi-LAT data. This inconsistency may stem from unmodeled nucleus-nucleus interactions at kinetic energies near or below $E_p^\mathrm{k} = 0.28$~GeV/nucleon, which requires investigation with dedicated hadronic interaction codes.

\item Our framework cannot explain the observed anti-correlation between the gamma-ray flux and the solar cycle. This limitation may arise from our treatment of Alfv\'{e}n wave turbulence calculation in only one single flux tube, which restricts the formation of long-wavelength turbulence. A future coupled-field model that describes energy transfer between adjacent flux tubes could generate the necessary long-wavelength Alfv\'{e}n wave turbulence to more effectively scatter higher-energy GCRs and pave the way for understanding this anti-correlation.
\end{enumerate}

\begin{acknowledgments}
We are grateful for valuable discussions with Ofer Cohen, Hugh Hudson, Mikhail Malkov, and Eleonora Puzzoni. This work was supported by NASA Grant No.\ {80NSSC25K7761}. {J.T.L.} and {A.H.G.P.} were additionally supported by NASA Grant Nos.\ {80NSSC20K1354} and {80NSSC22K0040}. {M.A.T.} was supported by NASA Grant No.\ {80NSSC23K0087} to the Smithsonian Astrophysical Observatory. {J.F.B.} was supported by National Science Foundation Grant No.\ {PHY-2310018}. Computational resources were provided by the Ohio Supercomputer Center. J.T.L. thanks the Harvard-Smithsonian Center for Astrophysics for their hospitality while parts of this manuscript were completed.
\end{acknowledgments}

\bibliographystyle{aasjournal}
\bibliography{reference}

\end{CJK*}
\end{document}